\documentclass[11pt,a4paper]{article}

\usepackage{Setting/jhepCAI190929}

\usepackage{ifpdf}

\usepackage{afterpage}
\usepackage[utf8]{inputenc}

\usepackage[T1]{fontenc} 

\usepackage{latexsym}
\usepackage{amsmath}
\usepackage{graphicx}
\usepackage{subfigure}
\usepackage{dcolumn}
\usepackage{bm}
\usepackage{amssymb}
\usepackage{latexsym}
\usepackage{datetime}
\usepackage{scrtime}
\usepackage{color}
\usepackage[dvipsnames, svgnames, x11names]{xcolor}

\usepackage{stmaryrd}
\usepackage{epsfig}
\usepackage{amsfonts}
\usepackage{amsmath}
\usepackage{graphicx}
\usepackage{color}
\usepackage{amssymb,amsmath,psfrag,slashed,graphicx}
\usepackage[normalem]{ulem}
\usepackage{bbm}

\def \nn {\nonumber}

\def\blue{\textcolor[rgb]{0,0,1}}

\allowdisplaybreaks

\def\be{\begin{equation}}
\def\ee{\end{equation}}
\def\ba{\begin{eqnarray}}
\def\ea{\end{eqnarray}}

\def\blue{\color{blue}}

\def\nn{\nonumber}
\def\lf{\left}
\def\rt{\right}

\begin{document}

\title{Perturbative unitarity and NEC violation in genesis cosmology}

\author[a,1]{Yong Cai,\note{Corresponding author: \texttt{\blue yongcai\_phy@outlook.com}}}
\author[a,2]{Ji Xu,\note{Corresponding author: \texttt{\blue xuji\_phy@zzu.edu.cn}}}
\author[b,c,3]{Shuai Zhao,\note{Corresponding author: \texttt{\blue szhao@odu.edu}}}
\author[d,4]{Siyi Zhou\note{Corresponding author: \texttt{\blue siyi@people.kobe-u.ac.jp}}}

\preprint{KOBE-COSMO-22-09, JLAB-THY-22-3668}

\affiliation[a]{School of Physics and Microelectronics, Zhengzhou University, Zhengzhou, Henan 450001, China}
\affiliation[b]{Department of Physics, Old Dominion University, Norfolk, Virginia 23529, USA}
\affiliation[c]{Theory Center, Thomas Jefferson National Accelerator Facility, Newport News, Virginia 23606, USA}
\affiliation[d]{Department of Physics, Kobe University, Kobe 657-8501, Japan}


\abstract{Explorations of the violation of null energy condition (NEC) in cosmology could enrich our understanding of the very early universe and the related gravity theories. Although a fully stable NEC violation can be realized in the ``beyond Horndeski'' theory, it remains an open question whether a violation of the NEC is allowed by some fundamental properties of UV-complete theories or the consistency requirements of effective field theory (EFT). We investigate the tree-level perturbative unitarity for stable NEC violations in the contexts of both Galileon and ``beyond Horndeski'' genesis cosmology, in which the universe is asymptotically Minkowskian in the past.
We find that the constraints of perturbative unitarity imply that we may need some unknown new physics below the cut-off scale of the EFT other than that represented by the ``beyond Horndeski'' operators.}


\maketitle


\section{Introduction}

Inflation \cite{Guth:1980zm,Starobinsky:1980te,Linde:1981mu,Albrecht:1982wi} has achieved great successes in simultaneously explaining several puzzles of the Big Bang cosmology. More significantly, inflation predicted a nearly scale-invariant power spectrum of the primordial scalar perturbations, which has been confirmed by observations of the cosmic microwave background temperature anisotropy \cite{Planck:2018vyg,Planck:2018jri}. However, an inflationary universe is geodesically incomplete in the past \cite{Borde:1993xh,Borde:2001nh}. Furthermore, the swampland conjecture \cite{Agrawal:2018own} and the trans-Planckian censorship conjecture \cite{Bedroya:2019tba} may also reinforce the inference that inflation is not the final story of the early universe \cite{Cai:2019hge,Li:2019ipk}.

Alternatives to or completions of the inflationary scenario generally involve violating the null energy condition (NEC),\footnote{In modified theories of gravity, the NEC may need to be replaced by a more general condition, i.e., the null convergence condition \cite{Tipler:1978zz}.} which is quite robust and is crucial to the proof of the Penrose's singularity theorem, see \cite{Rubakov:2014jja} for a review. Is it possible to realize a completely healthy NEC violation? What is the underlying physics required for a healthy NEC violation? Whether violations of the NEC did took place in the very early universe?
Considerable progress have been made in looking for answers to these questions in gravity and cosmology, especially in the study of nonsingular cosmology, including cosmological bounce \cite{Piao:2003zm,Piao:2004me,Piao:2005ag,Cai:2007qw,Cai:2008qw,Qiu:2011cy,Cai:2012va,Liu:2013kea,Qiu:2013eoa,Cai:2013kja,Koehn:2013upa,Battarra:2014tga,Wan:2015hya,Koehn:2015vvy,Qiu:2015nha,Nojiri:2016ygo,Banerjee:2016hom,Nandi:2019xag} and genesis \cite{Creminelli:2010ba,Liu:2011ns,Wang:2012bq,Liu:2012ww,Creminelli:2012my,Hinterbichler:2012fr,Hinterbichler:2012yn,Liu:2014tda,Pirtskhalava:2014esa,Nishi:2015pta,Kobayashi:2015gga,Cai:2016gjd,Nishi:2016ljg}  (see \cite{Piao:2003ty,Piao:2007sv} for studies on slow expansion), see also \cite{Dubovsky:2005xd,Creminelli:2006xe,Nicolis:2009qm,Rubakov:2013kaa,Elder:2013gya,Ijjas:2016pad}. Additionally, NEC violations could also occur during inflation and induce enhanced power spectrum of primordial gravitational waves (GWs) \cite{Cai:2020qpu,Cai:2022nqv}.

Nonetheless, challenges remain in constructing a consistent effective field theory (EFT) to violate the NEC.
The ``no-go'' theorem proved in \cite{Libanov:2016kfc,Kobayashi:2016xpl} indicates that pathological instabilities of perturbations appear in either the NEC-violating period or sooner or later in spatial flat nonsingular cosmology constructed by Horndeski theory \cite{Horndeski:1974wa,Deffayet:2011gz,Kobayashi:2011nu}, see also \cite{Easson:2011zy,Ijjas:2016tpn,Ijjas:2016vtq,Kolevatov:2016ppi,Akama:2017jsa,Dobre:2017pnt,Ageeva:2018lko,Ageeva:2020gti,Ageeva:2020buc}.
It is then demonstrated explicitly with the EFT method \cite{Cai:2016thi,Creminelli:2016zwa,Cai:2017tku,Cai:2017dyi,Kolevatov:2017voe} that fully stable NEC-violating nonsingular cosmological models can be constructed in ``beyond Horndeski'' theories \cite{Gleyzes:2014dya,Gleyzes:2014qga}, see \cite{Cai:2017dxl,Cai:2017pga,Mironov:2018oec,Qiu:2018nle,Ye:2019frg,Ye:2019sth,Mironov:2019qjt,Akama:2019qeh,Mironov:2019mye,Ilyas:2020qja,Ilyas:2020zcb,Zhu:2021whu,Zhu:2021ggm,Mironov:2022ffa} for later developments. Notably, the physics represented by higher derivative ``beyond Horndeski'' operators plays an essential role in realizing fully stable NEC-violating nonsingular cosmology. So far, it remains an open question whether some fundamental properties of UV-complete theories or the consistency requirements of EFT allow a violation of the NEC.


Studies of perturbative unitarity in cosmology may throw some light upon the unknown new physics related to the very early universe, see e.g.,  \cite{Nicolis:2009qm,Cannone:2014qna,deRham:2017avq,deRham:2017aoj,Grall:2020tqc,Goodhew:2020hob,Cespedes:2020xqq,Kim:2021pbr,Melville:2021lst,Brandenberger:2022pqo}.
It would be interesting to see what the constraints of perturbative unitarity can tell about those higher derivative operators which are essential in fully stable NEC-violating nonsingular cosmology, see also \cite{deRham:2017aoj} for the case of $P(X)$ theory in the context of bouncing cosmology.
However, applying the results of quantum field theory to cosmological background requires great care. Additionally, the calculation of amplitudes (even at tree-level) could be a formidable (though not impossible) task for NEC violations constructed by ``beyond Horndeski'' theories, since there are too many perturbative interacting terms.

In this paper, we investigate the tree-level perturbative unitarity of a stable NEC violation in the context of Galilean and ``beyond Horndeski'' genesis. The calculation of scattering amplitudes is carried out at sufficient past time so that the spacetime can be treated as asymptotical Minkowski. Consequently, the calculation can be greatly simplified due to the asymptotic behavior of the genesis solution.
Throughout this paper we adopt natural units $c=\hbar=1$ and have a metric signature $(-,+,+,+)$.


\section{Perturbative unitarity and NEC violation in Galileon genesis}\label{Sec:GalileonGenesis}

In this section and the next, we investigate perturbative unitarity for a stable NEC violation in the context of genesis cosmology. For simplicity, we start by considering a genesis model constructed by the cubic Galileon theory, which is only able to guarantee the stability of perturbations during the genesis phase. A genesis model which is fully stable throughout the entire history can be constructed with the ``beyond Horndeski'' higher derivative operators, which will be carried out in Sec. \ref{Sec:bHGenesis}. In the following, we will focus our discussion on the physics of the NEC-violating phase.

\subsection{Setup}

A stable cosmological genesis can be realized with the action (see e.g., \cite{Creminelli:2010ba,Creminelli:2012my,Nicolis:2009qm})
\ba
\label{genesis-action}
&\,& S=\int
d^4x\sqrt{-g}\lf({M_{\rm P}^2\over 2}R +\lambda_1 e^{2\phi/M}X+{\lambda_2\over M^4}X^2 + {\lambda_3\over M^3}X\Box\phi \rt)\,,
\ea
where $\phi$ is a scalar field with a dimension of mass, $X=\nabla_{\mu}\phi\nabla^\mu\phi$, $\Box
\phi=\nabla_{\mu}\nabla^{\mu}\phi$; $\lambda_{1}$, $\lambda_{2}$ and $\lambda_{3}$ are dimensionless constants, $M_{\rm P}$ is the reduced Planck mass, $M<M_{\rm P}$ is some energy scale.
We will work with the flat Friedmann-Robertson-Walker metric,  i.e.,
\be \label{FRW-1}
ds^2=-dt^2+a^2(t)d\vec{x}^2\,.
\ee

The background equations can be given as
\ba
\label{eqH}
3 H^2
M_{\rm P}^2 &=& -\lambda_1 e^{2\phi/M}\dot{\phi }^2
+{3 \lambda_{2} \over M^4}\dot{\phi }^4
+{6 \lambda_{3} \over M^3} H \dot{\phi }^3
\,,
\\
\dot{H} M_{\rm P}^2 &=&
\lambda_1 e^{2\phi/M} \dot{\phi }^2
-{2 \lambda_{2} \over M^4}\dot{\phi }^4
-{3 \lambda_{3} \over M^3} H \dot{\phi}^3
+{\lambda_{3} \over M^3}\dot{\phi }^2 \ddot{\phi}
\,,  \label{dotH}
\\
0 &=&
\lambda_1 e^{2\phi/M}\lf( \ddot{\phi}
+3 H\dot{\phi }
+\frac{1}{M} \dot{\phi }^2 \rt)
-{6 \lambda_{2} \over M^4}\lf(H\dot{\phi }^3+\dot{\phi }^2 \ddot{\phi} \rt)
\nn\\&\,&
-{3\lambda_{3} \over M^3}\lf(3H^2 \dot{\phi}^2+\dot{H} \dot{\phi }^2+2 H \dot{\phi } \ddot{\phi}  \rt)
\,, \label{eomphi}
\ea
where only two of Eqs. (\ref{eqH}) to (\ref{eomphi}) are independent.

In the unitary gauge, the action (\ref{genesis-action}) can be mapped to the EFT action (\ref{action01}) (see Appendix. \ref{sec:app-EFT-1}), where those non-zero functions are
\ba &\,& f=1\,,\label{eq:f}
\\&\,&
\Lambda(t)= {\lambda_{2} \over M^4}\dot{\phi }^4 + {\lambda_{3} \over M^3}\dot{\phi }^2 (\ddot{\phi}+3H\dot{\phi})
 \,,
\\&\,&
c(t)= -\lambda_1 e^{2\phi/M}\dot{\phi }^2
+{2\lambda_{2} \over M^4}\dot{\phi }^4
- {\lambda_{3} \over M^3}\dot{\phi }^2 (\ddot{\phi}-3H\dot{\phi})
 \,,
\\&\,&
M_2^4(t)={2\lambda_{2} \over M^4}\dot{\phi }^4+{\lambda_{3} \over 2 M^3}\dot{\phi }^2 (\ddot{\phi}+3H\dot{\phi})
 \,,
\\&\,&
m_3^3(t)= {2\lambda_{3} \over M^3}\dot{\phi }^3 \,,  \label{tildem4}
\ea
up to quadratic order.

We will set $h_{i j}=a^{2} \mathrm{e}^{2 \zeta}\left(\mathrm{e}^{\gamma}\right)_{i j}$ and $\gamma_{i i}=0=\partial_{i} \gamma_{i j}$ in the unitary gauge.
The quadratic action of tensor perturbation is
\be
S_{\gamma}^{(2)}=\frac{M_{\rm P}^{2}}{8} \int d^{3}x d t\, a^{3} \left[\dot{\gamma}_{i j}^{2}- \frac{\left(\partial_{k} \gamma_{i j}\right)^{2}}{a^{2}}\right]\,,
\ee
which is same as that in general relativity.
The quadratic action of curvature perturbation in the unitary gauge can be written as
\be
\label{eft_action02}
S^{(2)}_\zeta=\int d^4xa^3 Q_s\lf[
\dot{\zeta}^2-c_s^2{(\partial \zeta)^2\over a^2} \rt]\,,
\ee
where
\ba Q_s&=&{1\over \gamma^2}\lf[-\dot{H} M_{\rm P}^2+\frac{4 \lambda _2 \dot{\phi }^4}{M^4}+\frac{3 \lambda _3^2 \dot{\phi }^6}{M^6 M_{\rm P}^2}+\frac{\lambda _3 \dot{\phi }^2 }{M^3}\left(\ddot{\phi}+3 H \dot{\phi } \right) \rt]\,,
\\
c_s^2&=& 1+{4 \lambda _2 M^{-1} \dot{\phi }^4  + 2 \lambda _3 \dot{\phi }^2 ( H  \dot{\phi } +2 \lambda _3 M_{\rm P}^{-2} M^{-3} \dot{\phi }^4- \ddot{\phi} )
\over 	
M_{\rm P}^2 M^3  \dot{H} -4 \lambda _2 M^{-1} \dot{\phi }^4
-\lambda _3 \dot{\phi }^2 (3 H\dot{\phi }  +3 \lambda _3  M_{\rm P}^{-2} M^{-3} \dot{\phi }^4+
	\ddot{\phi} )  }\,,
\ea
and $\gamma=H-\frac{\lambda _3 \dot{\phi }^3}{M^3 M_{\rm P}^2}$, see e.g., \cite{Cai:2016thi} for details.
In order to avoid the ghost and gradient instabilities of the scalar perturbations, we should have $Q_s>0$ and $c_s^2>0$.

Since $\gamma\neq H$ for a nonzero $\lambda_3$ in general, the region where pathological instabilities appear does not necessarily overlap with the region of NEC-violation \cite{Easson:2011zy}. Therefore, it is possible to obtain a stable NEC-violation with action (\ref{genesis-action}) by removing the instabilities of perturbations to the later NEC-preserving phase. These instabilities can be eliminated by ``beyond Horndeski'' higher derivative operators. However, we will not go into the details of curing these instabilities thoroughly in this section for simplicity, since we focus only on the physics of the NEC-violating phase (i.e., the genesis phase).

In order to apply the bounds of perturbative unitarity, it is more convenient to work with the spatial flat gauge, in which $\zeta=0$, $\phi(t,{\bf x})=\phi_0(t)+\delta\phi(t,{\bf x})$. For convenience, we define $\sigma(t,{\bf x})\equiv \delta\phi(t,{\bf x})$. The quadratic, cubic and quartic actions of $\sigma$ can be given as
\ba
S^{(2)}_\sigma&=& \int d^4x \sqrt{-g} \Big[\lambda _1 e^{\frac{2 \phi _0}{M}}\lf(\partial_\mu\sigma\partial^\mu\sigma + {4\over M}\partial_\mu\phi_0 \partial^\mu \sigma \sigma +  {2\over M^2}\partial_\mu \phi_0 \partial^\mu \phi_0 \sigma^2 \rt)
\nn\\
&\,&
\qquad\qquad\,\,\, + \frac{2 \lambda _2}{M^4}\lf(\partial_\mu \phi_0 \partial^\mu \phi_0 \partial_\nu\sigma\partial^\nu\sigma + 2 \partial_\mu \phi_0 \partial_\nu \phi_0 \partial^\mu\sigma \partial^\nu\sigma  \rt)
\nn\\
&\,&
\qquad\qquad\,\,\, + \frac{2 \lambda _3}{M^3} \lf(\Box\phi_0 \partial_\mu\sigma\partial^\mu\sigma - \nabla_\mu \nabla_\nu\phi_0 \partial^\mu\sigma\partial^\nu\sigma \rt)    \Big] \label{eq:s2-220721}
\,,
\\
%
S^{(3)}_\sigma&=& \int d^4x\sqrt{-g}\Big[\frac{2\lambda _1 }{M} e^{\frac{2 \phi _0}{M}} \lf( \sigma  \partial_\mu\sigma\partial^\mu\sigma + \frac{2}{M } \partial_\mu\phi_0 \partial^\mu \sigma \sigma^2
+ \frac{2}{3 M^2}\partial_\mu\phi_0\partial^\mu\phi_0 \sigma ^3 \rt)
\nn\\
&\,&\qquad\qquad\,\,\,
+ \frac{4\lambda _2}{M^4}\partial_\mu{\phi }_0 \partial^\mu{\sigma } \partial_\nu\sigma\partial^\nu\sigma
+ \frac{\lambda _3}{M^3}\partial_\mu\sigma\partial^\mu\sigma \Box\sigma
\Big]\,,
\\
%
S^{(4)}_\sigma &=& \int d^4x\sqrt{-g}\Big[\frac{2\lambda _1 }{M^2} e^{\frac{2 \phi _0}{M}} \lf( \sigma ^2
\partial_\mu\sigma\partial^\mu\sigma
+\frac{4}{3 M}\partial_\mu\phi_0 \partial^\mu{\sigma } \sigma^3
+ \frac{1}{3 M^2}\partial_\mu\phi_0 \partial^\mu \phi_0 \sigma ^4 \rt)
\nn\\
&\,&\qquad\qquad\,\,\,
+  \frac{\lambda _2}{M^4}\lf( \partial_\mu\sigma\partial^\mu\sigma \rt)^2
\Big]\,.
\ea

\subsection{A solution of Galileon genesis}\label{sec:GalileonGenesis-1}

For a genesis background (i.e., $H\simeq 0$), Eq. (\ref{eqH}) suggests
\be
\label{e2phi}
e^{2\phi/M}={3\lambda_2\over \lambda_1}{\dot{\phi}^2\over M^4}\,.
\ee
The solution is
\be \label{dotphi}  \dot{\phi}={M\over (-t)}\,,\quad t<0\,.
\ee
We have $\phi(t)=-M\ln(-Mt)+{M\over2}\ln{3\lambda_2\over \lambda_1}$ so that Eq. (\ref{e2phi}) is satisfied, which is valid in the regime $|t|\gg 1/M$.
With Eq. (\ref{dotH}), we find $\dot{H}={\lambda_2+\lambda_3\over M_{\rm P}^2 (-t)^4}$. Genesis requires the violation of the NEC, which indicates $\dot{H}>0$. Therefore, the condition
\be
\lambda_2+\lambda_3>0 \label{cond-lambda-1}
\ee
should be satisfied.

Obviously, the Hubble parameter
\be \label{HH}
H={\lambda_2+\lambda_3\over 3 M_{\rm P}^2 }{1\over (-t)^3} + C\,,
\ee
where the constant $C$ should be set as $0$ for genesis. A non-zero $C$ can be used for realizing NEC-violating inflation, see e.g.,  \cite{Cai:2020qpu,Cai:2022nqv}.
From Eq. (\ref{HH}), we have
\be a(t)=e^{\int H
	dt}=\exp\lf[{\lambda_2+\lambda_3\over 6 M_{\rm P}^2 }{1\over (-t)^2}\rt]\simeq
1+{\lambda_2+\lambda_3\over 6 M_{\rm P}^2 }{1\over (-t)^2}  \label{Gat}
\ee
for $|t|\gg 1/M$ while we have set $a(-\infty)=1$. As can be seen from Eq. (\ref{FRW-1}), the universe asymptotically tends to the Minkowski space in infinite past.

With Eqs. (\ref{e2phi}) to (\ref{HH}), we find in the unitary gauge that
\ba Q_s&\approx& \frac{27 \lambda _2  M_{\rm P}^4}{\left(2 \lambda
	_3 - \lambda _2\right)^2}(-t)^2 \,,
\\
c_s^2&\approx& \frac{2 \lambda _3-\lambda _2}{3 \lambda _2}\,,\label{eq:cs2bg}
\ea
where we have kept only the leading order terms in both $Q_s$ and $c_s^2$.
Therefore, we should have
\be
0<\lambda_2<2\lambda_3 \label{cond-lambda-2}
\ee
so that $Q_s>0$ and $c_s^2>0$ when $|t|\gg 1/M$. It should be pointed out that $\gamma={\lambda_2-2\lambda_3\over 3M_{\rm P}^2 (-t)^3}<0$ during the genesis phase under the condition (\ref{cond-lambda-2}). If we assume that the genesis phase eventually enters the standard hot Big Bang expansion, the $\gamma-$crossing problem would be inevitable. Therefore, instabilities of the scalar perturbations cannot be eliminated from the entire history of the universe for the action (\ref{genesis-action}). These instabilities are assumed to be cured by physics (e.g., the higher order ``beyond Horndeski'' operators \cite{Cai:2016thi,Creminelli:2016zwa,Cai:2017tku}) outside the genesis phase. However, we will focus only on a stable NEC-violating genesis phase in the following for our purpose.
Additionally, in order to avoid superluminal propagation of the scalar perturbations, we should have $c_s^2\leq 1$, i.e.,
\be
\lambda_3\leq 2\lambda_2\,.
\ee

In the spatial flat gauge, by using Eqs. (\ref{e2phi}) to (\ref{Gat}), we find
\ba
S^{(2)}_\sigma&=& \int d^4x \Big[ {\cal A}(t)\dot{\sigma}^2 -{\cal B}(t)(\partial_i \sigma)^2
\Big]\,, \label{S20519}
\ea
where
\be
{\cal A}= {3\lambda_2\over M^2(-t)^2}+{\cal O}\lf({1\over M^3(-t)^3} \rt)\,,\quad {\cal B}= {2\lambda_3-\lambda_2\over M^2(-t)^2 }+{\cal O}\lf({1\over M^3(-t)^3}
\rt) \label{eq:AB}
\ee
for $|t|\gg M^{-1}$. The mass term appears in the fist line of Eq. (\ref{eq:s2-220721}) can be safely disregarded since it implies $m^2\sim M^{-2}|t|^{-4}$.
Apparently, the sound speed squared in the spatial flat gauge can be given as $c_s^2={\cal B}/{\cal A} \approx (2\lambda_3-\lambda_2)/(3\lambda_2)$, which is consistent with Eq. (\ref{eq:cs2bg}). Since $\zeta\simeq H\delta\rho/\dot{\rho}\sim \sigma (-t)^{-2}$, Eq. (\ref{S20519}) is also consistent with Eq.  (\ref{eft_action02}).

Similarly, in the regime $|t|\gg M^{-1}$, we can obtain
\ba
&\,& S^{(3)}_\sigma= \int d^4x \lf[\frac{\lambda _3 }{M^3}\partial_\rho\sigma\partial^\rho\sigma g^{\mu \nu } \partial_\mu\partial_\nu\sigma
+{\cal O}\lf( {\sigma^3 \over M^3(-t) }\rt)
\rt]\,, \label{eq:S3}
\\
&\,&S^{(4)}_\sigma= \int d^4x\lf[
\frac{\lambda _2}{M^4}\lf( \partial_\mu\sigma\partial^\mu\sigma \rt)^2
+{\cal O}\lf( {\sigma^4 \over M^4(-t)^2}\rt)
\rt]\,.\label{eq:S4}
\ea
In the following, we will consider only the leading order terms in Eq. (\ref{eq:S3}) and (\ref{eq:S4}) for simplicity.
Consequently, the asymptotic behavior of genesis solution is able to greatly simplify the calculation of scattering amplitudes.

According to the ``no-go'' theorem \cite{Libanov:2016kfc,Kobayashi:2016xpl}, requiring the absence of instabilities {\it in the entire history} of a nonsingular Universe constructed by Horndeski theory generally indicates the strong coupling issue.
It seems that the strong coupling issue appears in the limit $t\rightarrow -\infty$, as can be seen in Eqs. (\ref{S20519}) and (\ref{eq:AB}). However, this issue is not physical since it disappears in the unitary gauge as $Q_s\sim (-t)^2$ in Eq. (\ref{eft_action02}).
This is because we require the absence of instabilities {\it only in the genesis phase}.
Whether such an issue is problematic even in the spatial flat gauge may require further investigation, see \cite{Ageeva:2018lko,Ageeva:2020gti,Ageeva:2020buc} for recent studies.
Additionally, the consistency of the perturbative expansion implies constraints on $\sigma$ or its derivatives, such as $|\partial \sigma/M|\ll (-t)^{-1}$ and $|\partial^2 \sigma/M|\ll (-t)^{-2}$, since the coefficients ${\cal A}$ and ${\cal B}$ of the quadratic action (\ref{S20519}) are suppressed by $|t|^2$.
As is well-known, the genesis solution is an attractor, namely, $\sigma$ decays as time goes on, see e.g., \cite{Creminelli:2010ba}.
Therefore, we will assume that we are working in a regime which is free from the above strong coupling issue and the inconsistency of perturbation theory.

Additionally, for simplicity, we set ${\cal A}\equiv {\cal B}$ so that $c_s^2\equiv 1$, which indicates $\lambda_3=2\lambda_2$. We define
\be
\tilde{\sigma}={\cal A}^{1/2}\sigma\,, \qquad \tilde{M}=M{\cal A}^{1/2}\approx \sqrt{3\lambda_2}/(-t)\,, \label{eq:tihuan-1}
\ee
such that
\ba
&\,& S^{(2)}_{\tilde{\sigma}}= \int d^4x \lf[\dot{\tilde{\sigma}}^2 - (\partial_i \tilde{\sigma})^2 \rt]\,,\label{eq:S2-2}
\\
&\,& S^{(3)}_{\tilde{\sigma}}= \int d^4x \lf[ \frac{\lambda _3 }{\tilde{M}^3} \partial_\rho{\tilde{\sigma}}\partial^\rho{\tilde{\sigma}} g^{\mu \nu } \partial_\mu\partial_\nu{\tilde{\sigma}}
\rt]\,, \label{eq:S3-2}
\\
&\,&S^{(4)}_{\tilde{\sigma}}= \int d^4x\lf[
\frac{\lambda _2}{\tilde{M}^4} \lf( \partial_\mu{\tilde{\sigma}}\partial^\mu{\tilde{\sigma}} \rt)^2
\rt]\,,\label{eq:S4-2}
\ea
where we have neglected higher order terms. From Eq. (\ref{eq:S2-2}), we find that the dispersion relation is simply $F=\omega^2-k^2=0$.
Apparently, the interactions of $\tilde{\sigma}$ depend on $\tilde{M}$, or equivalently, on time $t$. In the regime $|t|\gg 1/M$, the time scale of the scattering processes we consider will be $\Delta t\ll |t|$. Therefore, we can approximately treat $\tilde{M}$ as a constant in the calculation of the scattering amplitudes.
It should be pointed that $\Delta t\ll |t|$ indicates the energy scale of the scattering process satisfies $M> E\sim 1/\Delta t \gg E_{\rm IR} \simeq |t|^{-1}$, where $M$ is the UV cut-off scale of the EFT (\ref{genesis-action}) and $E_{\rm IR}$ can be treated as an IR cut-off scale of these scattering processes we consider. The earlier era we go into the genesis phase, the smaller $E_{\rm IR}$ we get.



\subsection{Perturbative unitarity}\label{sec:PU-220717-1}

Some bounds on scattering amplitudes can be established in terms of optical theorem, which is a straightforward consequence of the unitarity of the ${\cal S}$-matrix, i.e., ${\cal S} {\cal S}^\dag =\mathbbm{1}$. Inserting ${\cal S}=\mathbbm{1}+i{\cal T}$, we have
\begin{eqnarray}\label{nunitarity_T}
-i({\cal T}-{\cal T}^\dag)={\cal T} {\cal T}^\dag \,.
\end{eqnarray}
Let us take the matrix element of this equation between initial states $| A \rangle$ and final states $| B \rangle$. Then express the ${\cal T}$-matrix elements as invariant matrix elements $\mathcal{M}$ times 4-momentum-conserving delta functions, where
\begin{align}
 \langle B | \mathcal T | A \rangle = (2\pi)^4 \delta^4 (p_A-p_B) {\cal M}(A \rightarrow B)\,.
\end{align}
Eq. (\ref{nunitarity_T}) becomes
\begin{eqnarray}
&&-i\left[\mathcal{M}\left(A \rightarrow B\right)-\mathcal{M}^{*}\left(B \rightarrow A\right)\right] \nn\\
=&&\sum_{m}\left(\prod_{i=1}^{m} \int \frac{d^{3} q_{i}}{(2 \pi)^{3}} \frac{1}{2 E_{i}}\right) \mathcal{M}^{*} \left(A \rightarrow\left\{q_{i}\right\}\right) \mathcal{M}\left(B \rightarrow\left\{q_{i}\right\}\right) \times(2 \pi)^{4} \delta^{(4)}(P_B-\sum_{i} q_{i}) \,, \nn\\
\end{eqnarray}
where $\{q_i\}$ is the inserted complete set of intermediate states. We are interested in the magnitude of scattering amplitude $\mathcal{M}\left(A \rightarrow\left\{q_{i}\right\}\right)$ with a typical high energy scale $E$, thus we take $A=B$,
\begin{eqnarray}\label{Im_bound}
2 \textrm{Im} \mathcal{M}\left(A \rightarrow A\right) = \sum_{m}\left(\prod_{i=1}^{m} \int \frac{d^{3} q_{i}}{(2 \pi)^{3}} \frac{1}{2 E_{i}}\right) (2 \pi)^{4} \delta^{(4)}(P_A-\sum_{i} q_{i}) \left|\mathcal{M}\left(A \rightarrow\left\{q_{i}\right\}\right)\right|^2 \,.
\end{eqnarray}

The mass dimensions of an $n$-particle scattering amplitude is
\begin{eqnarray}\label{bound_1}
[\mathcal{M}]=4-n \,,
\end{eqnarray}
where $n$ is the total number of particles involved in the process, i.e., the in+out particles. According to Eq.(\ref{bound_1}), a natural bound can be established on $\textrm{Im}\mathcal{M}\left(A \rightarrow A \right)$,
\begin{eqnarray}
\textrm{Im} \mathcal{M}\left(A \rightarrow A\right) \leq \left| \mathcal{M}\left(A \rightarrow A\right) \right| \leq E^{4-2n_A} \,.
\end{eqnarray}
Substituting this bound back into Eq.(\ref{Im_bound}), we have
\begin{eqnarray}\label{constrain}
\left|\mathcal{M}\left(A \rightarrow\left\{q_{i}\right\}\right)\right| \leq E^{4-(n_A+n_{q_i})} \,.
\end{eqnarray}

If weakly coupled UV completion was respected, this bound condition have to be satisfied by tree-level amplitudes. However, the requirement of UV completion may be too stringent for the EFT of a stable NEC-violating nonsingular cosmology.
In the regime of EFT, the constraint of perturbative unitarity can be given as \cite{deRham:2017aoj}
\be
\lf|\mathcal{M}_{4,\ell}\rt|\leq 8\pi^2\,.\label{constrain-EFT-1}
\ee
For four point scattering, where $n_A=n_{q_i}=2$, the constraint can be translated into the following partial wave representation
\begin{align}
 \mathcal{M}_{4, \ell}(\mathsf{s})=\int_{-1}^{1} \mathrm{~d} \cos \theta P_{\ell}(\cos \theta) \mathcal{M}_{4}(\mathsf{s}, \theta) \,,
\end{align}
where $\mathsf{s}$ is the Mandelstam variable, the subscript $4$ denotes the number of external legs.
As for the $2N$-point functions, we will have the constraint $\lf|\mathcal{M}_{2N}\rt|\lesssim \mathsf{s}^{2-N}$. {Specifically, we will consider the five-point scattering, for which the constraint is $\lf|\mathcal{M}_{5}\rt|\lesssim \mathsf{s}^{-1/2}$.}

\subsubsection{Constraints from four-point scattering}\label{sec:4GPUB-1}

\begin{figure}[htbp]
	\begin{center}
		\includegraphics[width=4.0in]{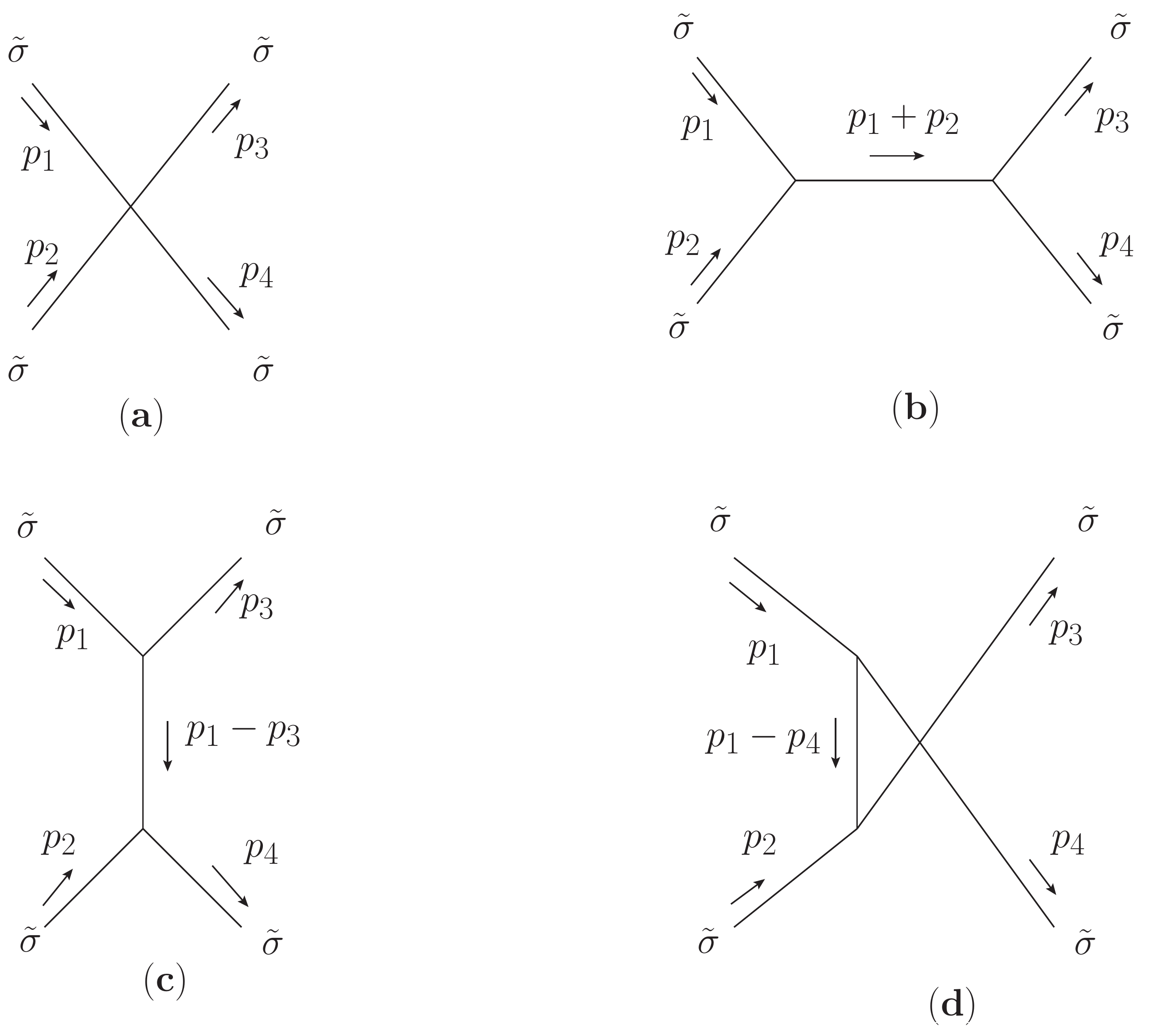}
	\end{center}
	\caption{Feynman diagrams for the four-point scattering $\tilde{\sigma}\tilde{\sigma}\to\tilde{\sigma}\tilde{\sigma}$.}
	\label{four_point}
\end{figure}

We will calculate the amplitude of four-point scattering $\tilde{\sigma}\tilde{\sigma}\to\tilde{\sigma}\tilde{\sigma}$ to find out what the constraints of perturbative unitarity, i.e., (\ref{constrain}) and (\ref{constrain-EFT-1}), can tell for a stable NEC-violating Galileon genesis.
From the actions (\ref{eq:S2-2}), (\ref{eq:S3-2}) and (\ref{eq:S4-2}), we can deduce the Feynman rules for $3\tilde{\sigma}$ and $4\tilde{\sigma}$ vertexes, which are collected in Fig. \ref{four_point}.
\begin{figure}[htbp]
	\begin{center}
		\includegraphics[width=6.0in]{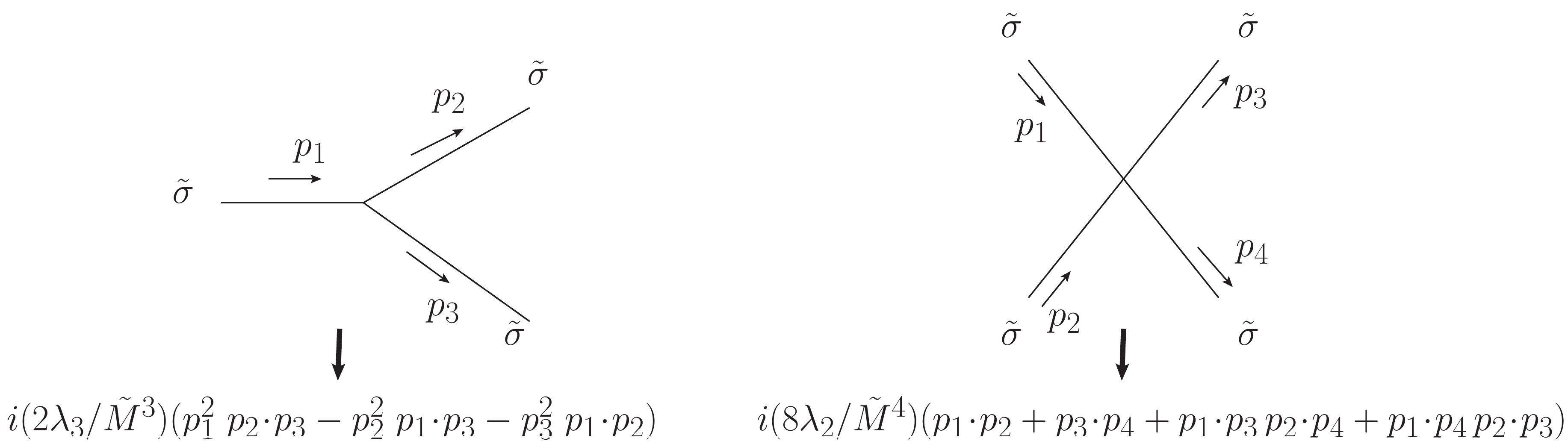}
	\end{center}
	\caption{Feynman rules for $3\tilde{\sigma}$ and $4\tilde{\sigma}$ vertexes.}
	\label{FeynmanRule}
\end{figure}

Having Feynman rules in hand, the amplitudes for every diagram which is shown in Fig. \ref{four_point} can be deduced,
\begin{eqnarray}
i \mathcal{M}^{(a)}(\tilde{\sigma}\tilde{\sigma}\to\tilde{\sigma}\tilde{\sigma}) &=& i\frac{8\lambda_2}{\tilde{M}^4} \left( p_1 \cdot p_2 ~ p_3 \cdot p_4 + p_1 \cdot p_3 ~ p_2 \cdot p_4 + p_1 \cdot p_4 ~ p_2 ~\cdot p_3 \right) \,,\nn\\
i \mathcal{M}^{(b)}(\tilde{\sigma}\tilde{\sigma}\to\tilde{\sigma}\tilde{\sigma}) &=& \frac{i2\lambda_3}{\tilde{M}^3}\left( -p_1^2 ~ p_2\cdot(p_1+p_2) -p_2^2 ~ p_1\cdot(p_1+p_2) + (p_1+p_2)^2 p_1 \cdot p_2 \right) \frac{i}{-(p_1+p_2)^2} \nn\\
&&\times \frac{i2\lambda_3}{\tilde{M}^3}\left( (p_1+p_2)^2 ~ p_3\cdot p_4 -p_3^2 ~ (p_1+p_2)\cdot p_4 - p_4^2 ~ (p_1+p_2) \cdot p_3 \right) \,,\nn\\
i \mathcal{M}^{(c)}(\tilde{\sigma}\tilde{\sigma}\to\tilde{\sigma}\tilde{\sigma}) &=& \frac{i2\lambda_3}{\tilde{M}^3}\left( p_1^2 ~ p_3\cdot(p_1-p_3) -p_3^2 ~ p_1\cdot(p_1-p_3) - (p_1-p_3)^2 p_1 \cdot p_3 \right) \frac{i}{-(p_1-p_3)^2} \nn\\
&&\times \frac{i2\lambda_3}{\tilde{M}^3}\left( p_2^2 ~ (p_3-p_1)\cdot p_4 -(p_3-p_1)^2 p_2\cdot p_4 - p_4^2 ~ p_2\cdot(p_3-p_1) \right) \,,\nn\\
i \mathcal{M}^{(d)}(\tilde{\sigma}\tilde{\sigma}\to\tilde{\sigma}\tilde{\sigma}) &=& \frac{i2\lambda_3}{\tilde{M}^3}\left( p_1^2 ~ p_4\cdot(p_1-p_4) -p_4^2 ~ p_1\cdot(p_1-p_4) - (p_1-p_4)^2 p_1 \cdot p_4 \right) \frac{i}{-(p_1-p_4)^2} \nn\\
&&\times \frac{i2\lambda_3}{\tilde{M}^3}\left( p_2^2 ~ (p_4-p_1)\cdot p_3 - (p_4-p_1)^2 p_2\cdot p_3 - p_3^2 ~ p_2\cdot(p_4-p_1) \right) \,.
\end{eqnarray}

As a result, we have
\begin{eqnarray}
\mathcal{M}(\tilde{\sigma}\tilde{\sigma}\to\tilde{\sigma}\tilde{\sigma}) &=& \frac{2\lambda_2}{\tilde{M}^4}\Big( \mathsf{s}^2+\mathsf{t}^2+\mathsf{u}^2 \Big) - \frac{\lambda_3^2}{\tilde{M}^6}\Big( \mathsf{s}^3+\mathsf{t}^3+\mathsf{u}^3 \Big) \,,\label{eq:220927-1}
\end{eqnarray}
where the Mandelstam variables
\begin{eqnarray}
\mathsf{s}=-(p_1+p_2)^2=-(p_3+p_4)^2 \,,\nn\\
\mathsf{t}=-(p_3-p_1)^2=-(p_2-p_4)^2 \,,\nn\\
\mathsf{u}=-(p_4-p_1)^2=-(p_2-p_3)^2 \,.
\end{eqnarray}


%

Recalling Eq. (\ref{eq:tihuan-1}) and $\lambda_3=2\lambda_2$, we find without surprise that the requirement of UV completion, i.e., (\ref{constrain}), cannot be satisfied by the stable NEC violation in the context of cubic Galileon genesis (\ref{genesis-action}).
In the EFT regime, the unitarity bound (\ref{constrain-EFT-1}) implies
\be
\mathsf{s} \lesssim {\lambda_2^{1/2}\over |t|^2} \sim \lambda_2^{1/2}(M_{\rm P}^2 H)^{2/3} \ll M^2 \,, \label{PUResult-1}
\ee
where we have used Eq. (\ref{HH}) and $|t|\gg M^{-1}$, and neglected the constant coefficients. Namely, the perturbative unitarity is violated at a scale $\sqrt{\mathsf{s}} \sim |t|^{-1}\sim \tilde{M}\ll M$.
However, the existences of the IR cut-off scale of the scattering processes we calculated and the UV cut-off scale of the EFT have already required $M> \sqrt{\mathsf{s}}\gg E_{\rm IR}\simeq |t|^{-1}$, which implies that the bound of perturbative unitarity (\ref{PUResult-1}) cannot be satisfied.


{
Therefore, although a stable NEC violation can be realized with cubic Galileon theory in the context of genesis cosmology, the bound of tree-level perturbative unitarity indicates that new physics should enter the EFT even blow the cut-off scale $M$.
Furthermore, (\ref{PUResult-1}) implies that the earlier era we go into the genesis phase, the more urgent we may need new physics at a lower scale, though the cosmological background is asymptotically Minkowskian.}

Notably, Ref. \cite{Nicolis:2009qm} has carried out a similar analysis of the strong coupling scale indicated by unitarity for the low-energy forward $2\rightarrow 2$ scattering of the background field in the Galileon theory. The second line of the lagrangian (3.2) in \cite{Nicolis:2009qm} seems equivalent to Eq. (\ref{genesis-action}) provided $c_2=f^2\simeq \lambda_1 M^2$ and $c_3={f^2\over 3H_0^2}\simeq \lambda_2$. However, two scales (i.e., $f$ and $H_0$) rather than one (i.e., $M$) were involved in \cite{Nicolis:2009qm}. As a result, the cut-off scales of the EFT actions are different, which are read as $H_0$ in \cite{Nicolis:2009qm} and $M$ in our case, unless $f=H_0=M$.\footnote{Note that the background field $\pi$ is dimensionless in \cite{Nicolis:2009qm}.}
As for the derivation of the strong coupling scales, our results in Eqs. (\ref{eq:220927-1}) and (\ref{PUResult-1}) could be consistent with that of Ref. \cite{Nicolis:2009qm} provided we set $f=H_0=\tilde{M}$.\footnote{See Eqs. (3.3) to (3.5) of \cite{Nicolis:2009qm}. The comparison of interpretations of these results is a little tricky, since the parameter $c_3$ in the second line of Eq. (3.2) is not exactly equal to the $c_3$ in the third line of Eq. (3.2) in \cite{Nicolis:2009qm}, where the latter one should depend on $\tilde{\pi}^{-4}$ actually. However, $c_3\simeq f^2/H_0^2$ has been used in the derivations of Eqs. (3.3) to (3.5) in \cite{Nicolis:2009qm}. Consequently, the third line of Eq. (3.2) in \cite{Nicolis:2009qm} will be equivalent to our Eqs. (\ref{eq:S2-2}) to (\ref{eq:S4-2}) if $f=H_0=\tilde{M}$, which indicates that the strong coupling scales $\sim \tilde{M}$ in both cases. }
Therefore, the situations and results in the two papers are not completely equivalent, but they can confirm each other to some extent.
Furthermore, we will consider explicitly also the five-point scattering in Sec. \ref{sec:5GPUB-1}.

\subsubsection{Constraints from five-point scattering}\label{sec:5GPUB-1}

For the five-point scattering $\tilde{\sigma}\tilde{\sigma}\to\tilde{\sigma}\tilde{\sigma}\tilde{\sigma}$, Eq. (\ref{constrain}) becomes
\begin{eqnarray}\label{constrain_five}
\left|\mathcal{M}\left(\tilde{\sigma}\tilde{\sigma}\to\tilde{\sigma}\tilde{\sigma}\tilde{\sigma}  \right)\right| \leq E^{-1} \,.
\end{eqnarray}
Similarly, the scattering amplitude for this five-point process can be written down. Here we define a set of Mandelstam variable basis
\begin{eqnarray}
\mathsf{s}_{12} &=& -(p_1+p_2)^2 \,, \qquad \mathsf{s}_{13} = -(p_1-p_3)^2 \,, \qquad \mathsf{s}_{14} = -(p_1-p_4)^2 \,,\nn\\
\mathsf{s}_{23} &=& -(p_2-p_3)^2 \,, \qquad \mathsf{s}_{24} = -(p_2-p_4)^2 \,.
\end{eqnarray}
This amplitude can be written in terms of Mandelstam variables
\begin{eqnarray}
\mathcal{M}\left(\tilde{\sigma}\tilde{\sigma}\to\tilde{\sigma}\tilde{\sigma}\tilde{\sigma}  \right) = \frac{\lambda_3}{\tilde{M}^9} \Big[ 4\lambda_2 \tilde{M}^2 G_1(\mathsf{s}_{ij}) + \lambda_3^2 G_2(\mathsf{s}_{ij})  \Big] \,,
\end{eqnarray}
where
\begin{eqnarray}
G_1(\mathsf{s}_{ij}) &=& -\mathsf{s}_{14}\Big( \mathsf{s}_{23}(2\mathsf{s}_{13}+\mathsf{s}_{14}+\mathsf{s}_{23}) +\mathsf{s}_{12}(2\mathsf{s}_{23}+\mathsf{s}_{13}) \Big) \nn\\
&&- \mathsf{s}_{24}\Big( \mathsf{s}_{13}(2\mathsf{s}_{12}+2\mathsf{s}_{14}+\mathsf{s}_{13})+\mathsf{s}_{23}(2\mathsf{s}_{13}+2\mathsf{s}_{14}+\mathsf{s}_{12})\Big) -\mathsf{s}_{13}\mathsf{s}_{24}^2 \,,\nn\\
G_2(\mathsf{s}_{ij}) &=& \mathsf{s}_{23}\Big( 2 \mathsf{s}_{13}^3+\mathsf{s}_{14}(\mathsf{s}_{14}+\mathsf{s}_{23})^2+\mathsf{s}_{13}^2(3\mathsf{s}_{14}+4\mathsf{s}_{23})+\mathsf{s}_{13}(3\mathsf{s}_{14}^2+4\mathsf{s}_{14}\mathsf{s}_{23}+2\mathsf{s}_{23}^2) \Big) \nn\\
&&+\mathsf{s}_{24}\Big( \mathsf{s}_{13}^3+3\mathsf{s}_{13}^2\mathsf{s}_{14}+3\mathsf{s}_{13}\mathsf{s}_{14}^2+2\mathsf{s}_{14}^3+4\mathsf{s}_{23}(\mathsf{s}_{13}+\mathsf{s}_{14})^2+3\mathsf{s}_{23}^2(\mathsf{s}_{13}+\mathsf{s}_{14}) \Big) \nn\\
&&+\mathsf{s}_{24}^2\Big( 2\mathsf{s}_{13}^2+4\mathsf{s}_{13}\mathsf{s}_{14}+4\mathsf{s}_{14}^2+3\mathsf{s}_{23}(\mathsf{s}_{13}+\mathsf{s}_{14}) \Big) \nn\\
&&+\mathsf{s}_{24}^2(\mathsf{s}_{13}+2\mathsf{s}_{14})+\mathsf{s}_{12}^3(\mathsf{s}_{13}+\mathsf{s}_{14}+\mathsf{s}_{23}+\mathsf{s}_{24}) \nn\\
&&+2\mathsf{s}_{12}^2\Big( \mathsf{s}_{13}^2+\mathsf{s}_{14}^2+\mathsf{s}_{23}^2+\mathsf{s}_{13}(\mathsf{s}_{14}+3\mathsf{s}_{23})+\mathsf{s}_{24}(\mathsf{s}_{23}+3\mathsf{s}_{14})+\mathsf{s}_{24}^2 \Big) \nn\\
&&+\mathsf{s}_{13}\Big( \mathsf{s}_{12}^3+\mathsf{s}_{14}^3+(\mathsf{s}_{23}+\mathsf{s}_{24})^3+\mathsf{s}_{13}^2(3\mathsf{s}_{14}+7\mathsf{s}_{23}+2\mathsf{s}_{24})+\mathsf{s}_{14}^2(2\mathsf{s}_{23}+7\mathsf{s}_{24}) \nn\\
&&+\mathsf{s}_{14}(2\mathsf{s}_{23}^2+4\mathsf{s}_{23}\mathsf{s}_{24}+7\mathsf{s}_{24}^2)+\mathsf{s}_{13}(3\mathsf{s}_{14}^2+7\mathsf{s}_{23}^2+4\mathsf{s}_{23}\mathsf{s}_{24}+2\mathsf{s}_{24}^2+4\mathsf{s}_{14}\mathsf{s}_{23}+4\mathsf{s}_{14}\mathsf{s}_{24}) \Big) \,.\nn\\
\end{eqnarray}

\begin{figure}[htbp]
	\begin{center}
		\includegraphics[width=5.5in]{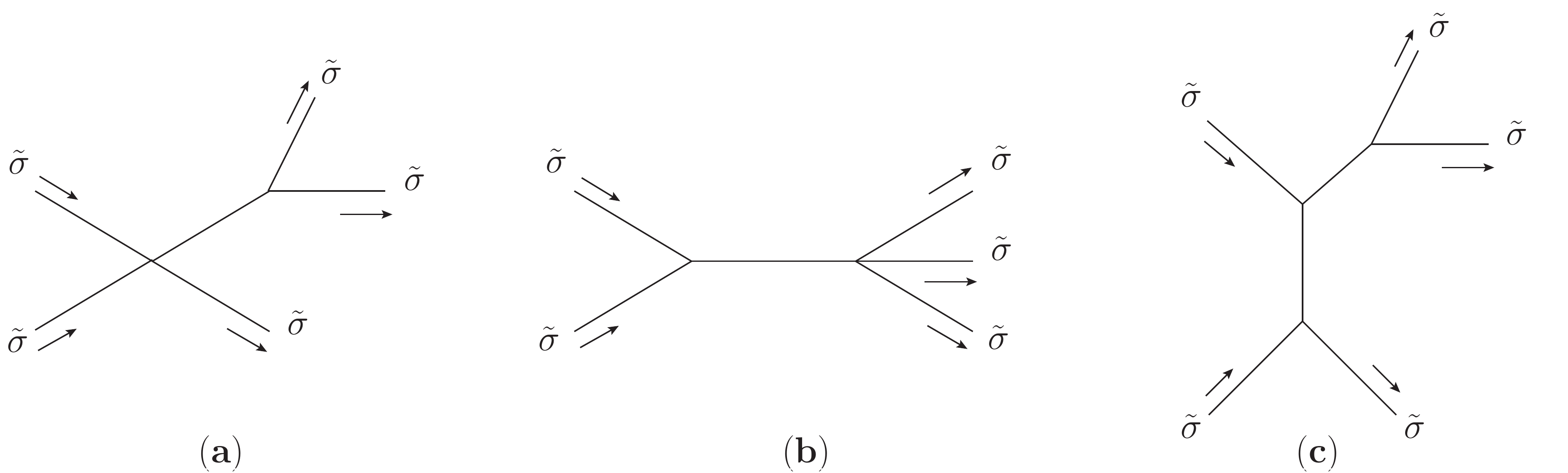}
	\end{center}
	\caption{Feynman diagrams for the five-point scattering $\tilde{\sigma}\tilde{\sigma}\to\tilde{\sigma}\tilde{\sigma}\tilde{\sigma}$. The topological inequivalent diagrams are presented and the momentum labels are omitted.}
	\label{five_point}
\end{figure}

One can resort to Fig. \ref{five_point} for relevant Feynman diagrams. { The unitarity bound Eq. (\ref{constrain_five}) again implies that the cubic Galileon genesis (\ref{genesis-action}) cannot be UV complete. In the EFT regime, the bound of perturbative unitarity $\lf|\mathcal{M}_{5}\rt|\lesssim \mathsf{s}^{-1/2}$ indicates that $\sqrt{\mathsf{s} }\lesssim |t|^{-1}\sim \tilde{M} \ll M$, which is consistent with (\ref{PUResult-1}), where we have assumed that $\mathsf{s}_{ij}\sim \mathsf{s}$ in our estimation. Therefore, the conclusion of Sec. \ref{sec:4GPUB-1} remains valid.}


\section{Perturbative unitarity and NEC violation in ``beyond Horndeski'' genesis}\label{Sec:bHGenesis}

The results of Secs. \ref{sec:4GPUB-1} and \ref{sec:5GPUB-1} motivate us to explore whether the new physics implied by the bounds of perturbative unitarity can be represented by ``beyond Horndeski'' operators, which are required by the absence of instabilities in the entire history of an NEC-violating nonsingular cosmology. In this section, we apply the constraints of perturbative unitarity to a stable NEC violation in the context of ``beyond Horndeski'' genesis.

\subsection{Setup}

It is discovered in Refs. \cite{Cai:2016thi,Creminelli:2016zwa,Cai:2017tku} that the EFT operator $\delta g^{00}R^{(3)}$ plays a crucial role in thoroughly eliminating the pathological instabilities of perturbations induced by a violation of the NEC while $m_4^2\neq {\tilde m}_4^2$ (i.e., $\alpha_{\mathrm{H}}\neq 0$), see Appendix. \ref{sec:app-EFT-1}. The covariant action
\begin{equation}
S=\int d^{4} x \sqrt{-g}\left[\frac{M_{\rm P}^{2}}{2} R+P(\phi, X)\right]+S_{\delta g^{00} R^{(3)}} \label{eq:R3g00-1}
\end{equation}
can be used to construct nonsingular cosmology which is fully stable in the entire history, where $S_{\delta g^{00} R^{(3)}}=\int d^{4} x \sqrt{-g} L_{\delta g^{00} R^{(3)}}$,
\ba \label{eq:220122L-1}
L_{\delta g^{00} R^{(3)}}&=& \frac{f_{1}(\phi)}{2} \delta g^{00} R^{(3)} \nn\\
&=& \frac{f}{2} R-\frac{X}{2} \int f_{\phi \phi} d \ln X-\left(f_{\phi}+\int \frac{f_{\phi}}{2} d \ln X\right) \square \phi \\
&\,&+\frac{f}{2 X}\left[\phi_{\mu \nu} \phi^{\mu \nu}-(\square \phi)^{2}\right]-\frac{f-2 X f_{X}}{X^{2}}\left[\phi^{\mu} \phi_{\mu \rho} \phi^{\rho \nu} \phi_{\nu}-(\square \phi) \phi^{\mu} \phi_{\mu \nu} \phi^{\nu}\right]\,,\nn
\\
f(\phi, X)&=& f_{1}\left(1+{X/f_{2}}\right)\,,\qquad \delta g^{00}= {X/\dot{\phi}^{2}(t)}+1= {X/f_{2}(t(\phi))}+1\,, \label{eq:220122fg-1}
\ea
$\phi_{\mu}=\nabla_{\mu} \phi$, $\phi^{\mu}=\nabla^{\mu} \phi$, $f_X=df/dX$, $f_\phi=df/d\phi$ and $f_{\phi\phi}=d^2f/d\phi^2$.
It is demonstrated in Refs. \cite{Cai:2017dyi,Kolevatov:2017voe} that the action (\ref{eq:R3g00-1}) belongs to the ``beyond Horndeski'' (GLPV) theory \cite{Gleyzes:2014dya,Gleyzes:2014qga}, which corresponds to $m_4^2\neq {\tilde m}_4^2$ in the EFT action (\ref{action01}).

The integral terms in Eq. (\ref{eq:220122L-1}) and the constraints (\ref{eq:220122fg-1}) guarantee that the background evolution and the quadratic action of tensor perturbation are unaffected by $S_{\delta g^{00} R^{(3)}}$.
However, we can relax such requirements so that we can get rid of the integral terms and the constraints (\ref{eq:220122fg-1}). Inspired by (\ref{eq:220122L-1}), we can construct a fully stable NEC-violating genesis model with the extended action
\ba
S&=&\int d^{4} x \sqrt{-g}\Big\{\frac{1}{2}[M_{\rm P}^{2}+ F(\phi,X)] R+P(\phi, X)+G(\phi, X)\Box\phi
\nn\\
&\,&
+ \frac{F}{2 X}\left[\phi_{\mu \nu} \phi^{\mu \nu}-(\square \phi)^{2}\right]-\frac{F-2 X F_{X}}{X^{2}}\left[\phi^{\mu} \phi_{\mu \rho} \phi^{\rho \nu} \phi_{\nu}-(\square \phi) \phi^{\mu} \phi_{\mu \nu} \phi^{\nu}\right]
\Big\} \label{eq:210726-1}
\,,
\ea
see \cite{Ilyas:2020qja,Ilyas:2020zcb,Zhu:2021whu,Zhu:2021ggm} for recent studies.
It can be checked that the action (\ref{eq:210726-1}) still belongs to the ``beyond Horndeski'' (GLPV) theory \cite{Gleyzes:2014dya,Gleyzes:2014qga},\footnote{See e.g. Eq. (4) of \cite{Langlois:2017dyl} or Eq. (2.17) of \cite{Langlois:2018dxi}, see also the footnote on page 2 of \cite{Zhu:2021ggm}. } which is free from the Ostrogradsky instabilities. In the special case of $F=2XF_X$, the action (\ref{eq:210726-1}) reduces to a subclass of Horndeski theory.

Expanding around a cosmological background, the quadratic order of action (\ref{eq:210726-1}) in perturbations can be written as (\ref{eq:21080102}), in which
\ba
&\,&{\cal M}^2=M_{\rm P}^2,\quad \alpha_{\rm T}=F/M_{\rm P}^2 \,, \nn\\
&\,&
\alpha_{\rm K}=(4 \dot{\phi }^4
P_{XX}
-2 \dot{\phi }^2 P_{X}
+12 H \dot{\phi }^3 G_{X}
-12 H \dot{\phi }^5 G_{XX}
+2 \dot{\phi }^2 G_{\phi }
-2\dot{\phi }^4 G_{\phi X})/(H^2M_{\rm P}^2) ,
\nn\\
&\,&
\alpha_{\rm B}= \frac{-2 \dot{\phi }^3 G_{X} -2 \dot{\phi } \ddot{\phi}F_{X}  + \dot{\phi } F_{\phi } }{2H M_{\rm P}^2} ,
\quad \alpha_{\rm H}= {F-2XF_X \over M_{\rm P}^2}
\nn\\
&\,&\alpha_{\rm L}=0, \quad \beta_1= \beta_2=\beta_3=0\,.
\ea
Since $\alpha_{\rm L}=\beta_i=0$, the action (\ref{eq:210726-1}) does not involve those characteristic degenerate higher-order scalar-tensor (DHOST) operators in (\ref{eq:210726-1}). The parameter $\alpha_{\rm H}$ vanishes (i.e., $m_4^2= {\tilde m}_4^2$) for Horndeski theory and becomes non-zero (i.e., $m_4^2\neq {\tilde m}_4^2$) for ``beyond Horndeski'' (GLPV) theory, as we have mentioned.


The background equations of (\ref{eq:210726-1}) can be obtained as
\ba
3 M_{\rm P}^{2} H^{2} &=& -P+6 H \dot{\phi}^{3} G_{X}+\dot{\phi}^{2} G_{\phi}-2 \dot{\phi}^{2} P_{X}
-3 H \dot{\phi} F_{\phi}+6 H \dot{\phi}^{3} F_{\phi X}\,, \label{eq:220123H}
\\
M_{\rm P}^2 \dot{H} &=& \dot{\phi }^2 P_{X}-\dot{\phi }^2 G_{\phi } -3 H \dot{\phi }^3 G_{X}+\dot{\phi }^2 \ddot{\phi} G_{X}
\nn\\ &\,&
+\frac{3}{2} H \dot{\phi} F_{\phi}-\frac{1}{2} F_{\phi} \ddot{\phi}-\frac{1}{2} \dot{\phi}^{2} F_{\phi \phi}+\dot{\phi}^{2} \ddot{\phi} F_{\phi X}-3 H \dot{\phi}^{3} F_{\phi X}\,.\label{eq:210805-01}
\ea
Obviously, $F(\phi,X)$ will not appear in the background equations when $F$ is independent on $\phi$.

If we write the action (\ref{eq:210726-1}) in terms of the operators in (\ref{action01}), we do not need to include the operators $\delta K \dot{\delta g^{00}}$, $(\dot{\delta g^{00}})^2$ or $(\partial_i{\delta g^{00}})^2$. The coefficients $c(t)$ and $\Lambda(t)$ can be determined by $c(t)={M_{\rm P}^2\over2}(\dot{f}H-2f\dot{H}-\ddot{f})$ and $\Lambda(t)={M_{\rm P}^2\over2}(5\dot{f}H+6f H^2+2f\dot{H}+\ddot{f})$, where $f=1+F/M_{\rm P}^2$, $H$ is solved by Eqs. (\ref{eq:220123H}) and (\ref{eq:210805-01}). The explicit expressions of $c(t)$, $\Lambda(t)$ and $M_2^4$ are clumsy, while $m_3^3=2\dot{\phi}^3 G_X$, $m_4^2=-F/2$ and $\tilde{m}_4^2=-XF_X$.

The quadratic action of tensor perturbation is
\be
S_{\gamma}^{(2)}=\frac{M_{\rm P}^{2}}{8} \int d^{3}x d t\, a^{3} Q_{T}\left[\dot{\gamma}_{i j}^{2}-c_{T}^{2} \frac{\left(\partial_{k} \gamma_{i j}\right)^{2}}{a^{2}}\right]\,,
\ee
where
\be
Q_T= 1,\quad c_T^2=1+\alpha_{\rm T} \,.\label{bH:eq:210805-02}
\ee
Apparently, for a non-vanishing $F(\phi,X)$, the propagating speed of primordial gravitational waves will be modified during the genesis phase, which could be able to generate interesting features in the power spectrum of primordial gravitational waves (\cite{Cai:2015yza,Cai:2016ldn,Wang:2016tbj,Cai:2022nqv}).
We require $-M_{\rm P}^2<F\leq0$ so that  $0< c_T^2 \leq 1$.


The quadratic action of curvature perturbation in the unitary gauge can be written as
\be
\label{bH:eft_action02}
S^{(2)}_\zeta=\int d^{3} x d t\, a^3 Q_s\lf[
\dot{\zeta}^2-c_s^2{(\partial \zeta)^2\over a^2} \rt]\,,
\ee
where
\ba Q_s&=& {M_{\rm P}^2\over 2}{\alpha_{\rm K}+6\alpha_{\rm B}^2 \over ( 1+\alpha_{\rm B})^{2}}\,,\label{bH:eq:210805-03}
\\
c_s^2&=& {M_{\rm P}^2\over Q_s}\lf\{{1\over a}{{\rm d}\over {\rm d}t}\lf[{a(1+\alpha_{\rm H})\over H (1+\alpha_{\rm B})}\rt] -(1+\alpha_{\rm T})\rt\} \,,\label{bH:eq:210805-04}
\ea
see e.g., \cite{Langlois:2017mxy}.
In order to avoid the ghost and gradient instabilities of $\zeta$, $Q_s>0$ and $c_s^2>0$ are required, respectively.
We also require $c_s^2\leq1$ so that there is no superluminal propagation of the scalar perturbation modes.
It is convenient to define $\gamma\equiv H(1+\alpha_{\rm B})$ and $Q_{\tilde{m}_4}\equiv 1+\alpha_H$, which are usually used in the analysis of these instability problems. The ways to overcome the ``no-go'' theorem have been discussed explicitly in \cite{Cai:2017tku} with these quantities $\gamma$ and $Q_{\tilde{m}_4}$.

\subsection{A solution of ``beyond Horndeski'' genesis}\label{bH:Sec:210802-01}

Based on the previous section, a model of fully stable genesis can be constructed. In this section, we will set the free functions $F(\phi,X)$, $P(\phi,X)$ and $G(\phi,X)$ in (\ref{eq:210726-1}) as
\ba
F(\phi,X)&=& M_{\rm P}^2\lf({\kappa_1\over M^4}X+ {\kappa_2 \over M^8}X^2 \rt) \label{eq:FX0715}
\,,
\\
P(\phi,X)&=& \lambda_1 e^{2\phi/M}X+{\lambda_2\over M^4}X^2 \,,
\\
G(\phi,X)&=&  {\lambda_3\over M^3}X \,,
\ea
where $\kappa_1$, $\kappa_2$, $\lambda_1$, $\lambda_2$, $\lambda_3$ are dimensionless constants.
Apparently, the action (\ref{eq:210726-1}) can be viewed as the addition of higher order ``beyond Horndeski'' operators to the action (\ref{genesis-action}) in Sec. \ref{Sec:GalileonGenesis}.
The background evolution of the universe will not be affected by $F$. As a result, the solution of background evolution of genesis will be exactly the same as that introduced in Sec. \ref{sec:GalileonGenesis-1}.
%

Substituting the genesis solution (i.e., Eqs. (\ref{dotphi}), (\ref{HH}) and (\ref{Gat})) into Eqs. (\ref{bH:eq:210805-02}), (\ref{bH:eq:210805-03}) and (\ref{bH:eq:210805-04}), we have
\ba
Q_T&=& 1,\quad c_T^2=1-\frac{\kappa _1}{M^2 (-t)^2} +\frac{\kappa _2}{M^4 (-t)^4}\,.
\\
Q_s&\approx& \frac{27 \lambda _2  M_{\rm P}^4}{\left(\lambda _2-2 \lambda
	_3 -3 \kappa _1 M_{\rm P}^2 M^{-2}\right)^2}(-t)^2 \,,
\\
c_s^2&\approx& -\frac{\lambda _2-2 \lambda _3-3\kappa_1 M_{\rm P}^2 M^{-2}}{3 \lambda _2} \label{eq:cs2-220716}
\ea
for $|t|\gg M^{-1}$.
We have kept only the leading order terms in $Q_s$ and $c_s^2$, where $\kappa_2$ does not appear. Apparently, we should require $\kappa_1>0$ so that $c_T\leq 1$. In order to guarantee that $Q_s>0$ and $0<c_s^2\leq1$, we should have $\lambda_2>0$, $\lambda _2-2 \lambda_3 -3 \kappa _1 M_{\rm P}^2 M^{-2}<0$ and $4\lambda _2-2 \lambda_3 -3 \kappa _1 M_{\rm P}^2 M^{-2}\geq 0$. Additionally, we may also consider adding a term $\sim (-X/M^4)^n$ in $F(X)$, where $n$ is a constant. However, we will find $c_s>1$ in the limit $|t|\gg M^{-1}$ for $n<1$.
Therefore, given the genesis solution and the requirement of $0<c_s^2\leq1$, the formulation of $F(X)$ in Eq. (\ref{eq:FX0715}) is quite general.

The quantities
\be
\gamma
=\frac{\lambda _2-2 \lambda _3 -3\kappa_1 M_{\rm P}^2 M^{-2}}{3M_{\rm P}^2 (-t)^3 }
+\frac{2 \kappa _2}{M^4 (-t)^5}\,,
\ee
\be
Q_{\tilde{m}_4} = 1+\frac{\kappa _1}{M^2 (-t)^2}-\frac{3 \kappa _2}{M^4 (-t)^4}\,.
\ee
Hence, $\gamma<0$ and $Q_{\tilde{m}_4}>0$ in the limit $|t|\gg M^{-1}$.
If we assume that the universe eventually enters the standard hot Big Bang expansion, during which $\gamma=H>0$, we will find $\gamma$ cross $0$ at some time $t_\gamma$ after the end of the genesis phase.\footnote{Note that the genesis solution is no longer applicable at $t_\gamma$.}
In order to avoid the gradient instability induced by the $\gamma-$crossing around $t_\gamma$, we should carefully design the behavior of $Q_{\tilde{m}_4}$, see e.g., the condition (13) of \cite{Cai:2017tku}, see also \cite{Zhu:2021ggm} for an example of the numerical simulation.
In this paper, instead of handling the explicit constructions of $\gamma$ and $Q_{\tilde{m}_4}$, we will focus our discussion on the behavior of genesis in the limit $|t|\gg M^{-1}$ for our purpose.

In the spatial flat gauge, we have
\ba
S^{(2)}_\sigma&=& \int d^4x \Big[ {\cal A}(t)\dot{\sigma}^2 -{\cal B}(t)(\partial_i \sigma)^2
\Big]\,, \label{S20714}
\ea
where
\be
{\cal A}= {3\lambda_2\over M^2(-t)^2}+{\cal O}\lf({1\over M^3(-t)^3} \rt)\,,\quad {\cal B}= {2\lambda_3-\lambda_2 + 3\kappa_1 M_{\rm P}^2 M^{-2} \over M^2(-t)^2 }+{\cal O}\lf({1\over M^3(-t)^3}
\rt) \label{eq:AB-220716}
\ee
for $|t|\gg M^{-1}$.
The sound speed squared in the spatial flat gauge can be given as $c_s^2={\cal B}/{\cal A} \approx (2\lambda_3-\lambda_2 + 3\kappa_1 M_{\rm P}^2 M^{-2})/(3\lambda_2)$, which is consistent with Eq. (\ref{eq:cs2-220716}).
The arguments given in Sec. \ref{sec:GalileonGenesis-1} remain applicable.

In the limit $|t|\gg M^{-1}$, the leading order terms in the cubic and quartic actions of $\sigma$ can be found as
\ba
&\,& S^{(3)}_\sigma= \int d^4x {2M_{\rm P}^2\kappa_1 (-t)\over M^5} \lf[ g^{\mu \nu} g^{\alpha\beta}\partial_ \nu\sigma \partial_\alpha \dot{\sigma}\partial_\mu\partial_\beta \sigma
-\partial_i \sigma \partial_i \dot{\sigma} g^{\mu \nu}\partial_\mu\partial_ \nu\sigma
+g^{\mu \nu}\partial_\mu\dot {\sigma} \partial_ \nu\dot {\sigma} \dot{\sigma}
+{\cal O}(-t)^{-1}
\rt]\,, \label{bH:eq:S3}\nn
\\
&\,&S^{(4)}_\sigma= \int d^4x \frac{ M_{\rm P}^2 \kappa _1 (-t)^2}{M^6}\Big[
g^{\mu \alpha }g^{\nu \beta }\partial _{\mu }\sigma \partial _{\nu
}\sigma \partial _{\alpha }\partial _{\beta }\sigma  g^{\rho \gamma
}\partial_\rho\partial_\gamma \sigma
-g^{\mu \alpha }g^{\rho \beta }g^{\nu \gamma }\partial _{\alpha }\sigma
\partial _{\nu }\sigma \partial _{\rho }\partial _{\mu }\sigma
\partial _{\gamma }\partial _{\beta }\sigma
\nn\\ &\,&\qquad\qquad\qquad\qquad\qquad\quad
-g^{\alpha\beta}\partial_{\alpha }\sigma \partial_{\beta } \sigma  g^{\mu \nu }\partial _{\mu }\dot{\sigma
}\partial _{\nu }\dot{\sigma }
-4 g^{\alpha \beta }g^{\mu \nu }\partial _{\nu }\sigma
\partial _{\alpha }\dot{\sigma }\partial _{\mu }\partial _{\beta
}\sigma \dot{\sigma }
\nn\\ &\,&\qquad\qquad\qquad\qquad\qquad\quad
+ 4 \dot{\sigma } \partial _i\sigma \partial _i\dot{\sigma } g^{\mu \nu
}\partial_\mu\partial_\nu \sigma
-4 \dot{\sigma }^2 g^{\mu \nu }\partial _{\mu }\dot{\sigma }\partial
_{\nu }\dot{\sigma }
\nn\\ &\,&\qquad\qquad\qquad\qquad\qquad\quad
+g^{\alpha\beta}\partial_{\alpha }\sigma \partial_{\beta } \sigma g^{\mu \nu }\partial_\mu\partial_\nu \sigma \ddot{\sigma
}
+{\cal O}(-t)^{-1}
\Big]\,.\label{bH:eq:S4}
\ea
Apparently, the interactions are distinctive from that of the cubic Galileon genesis.

Similar to Sec. \ref{sec:GalileonGenesis-1}, we set ${\cal A}\equiv {\cal B}$ so that $c_s^2\equiv 1$ for simplicity, which indicates $\lambda_3=2\lambda_2- {3 \tilde{\kappa_1}}/2$, where $\tilde{\kappa}_1\equiv \kappa_1 M_{\rm P}^2 M^{-2}$. We define
\be
\tilde{\sigma}={\cal A}^{1/2}\sigma\,, \qquad \tilde{M}=M{\cal A}^{1/2}\approx \sqrt{3\lambda_2}/(-t)\,, \label{eq:tihuan-2-220716}
\ee
such that
\ba
&\,&S^{(2)}_{\tilde{\sigma}}= \int d^4x \lf[\dot{\tilde{\sigma}}^2 - (\partial_i \tilde{\sigma})^2 \rt]\,,\label{eq:S2-2-220716}
\\
&\,&S^{(3)}_{\tilde{\sigma}}= \int d^4x {2\tilde{\kappa}_1 (-t) \over \tilde{M}^3 } \lf(
g^{\mu \nu} g^{\alpha\beta}\partial_ \nu\tilde{\sigma} \partial_\alpha \dot{\tilde{\sigma}}\partial_\mu\partial_\beta \tilde{\sigma}
-\partial_i \tilde{\sigma} \partial_i \dot{\tilde{\sigma}} g^{\mu \nu}\partial_\mu\partial_\nu\tilde{\sigma}
+g^{\mu \nu}\partial_\mu\dot{\tilde{\sigma}} \partial_ \nu\dot{\tilde{\sigma}} \dot{\tilde{\sigma}}
\rt)\,, \label{eq:S3-2-220716}
\\
&\,&S^{(4)}_{\tilde{\sigma}}= \int d^4x {\tilde{\kappa}_1 (-t)^2 \over \tilde{M}^4}\Big(
g^{\mu \alpha }g^{\nu \beta }\partial _{\mu }\tilde{\sigma} \partial _{\nu
}\tilde{\sigma} \partial _{\alpha }\partial _{\beta }\tilde{\sigma}  g^{\rho \gamma
}\partial_\rho\partial_\gamma \tilde{\sigma}
-g^{\mu \alpha }g^{\rho \beta }g^{\nu \gamma }\partial _{\alpha }\tilde{\sigma}
\partial _{\nu }\tilde{\sigma} \partial _{\rho }\partial _{\mu }\tilde{\sigma}
\partial _{\gamma }\partial _{\beta }\tilde{\sigma}
\nn\\ &\,&\qquad\qquad\qquad\qquad\quad
-g^{\alpha\beta}\partial_{\alpha }\tilde{\sigma} \partial_{\beta } \tilde{\sigma}  g^{\mu \nu }\partial _{\mu }\dot{\tilde{\sigma}
}\partial _{\nu }\dot{\tilde{\sigma} }
-4 g^{\alpha \beta }g^{\mu \nu }\partial _{\nu }\tilde{\sigma}
\partial _{\alpha }\dot{\tilde{\sigma} }\partial _{\mu }\partial _{\beta
}\tilde{\sigma} \dot{\tilde{\sigma} }
\nn\\ &\,&\qquad\qquad\qquad\qquad\quad
+ 4 \dot{\tilde{\sigma} } \partial _i\tilde{\sigma} \partial _i\dot{\tilde{\sigma} } g^{\mu \nu
}\partial_\mu\partial_\nu \tilde{\sigma}
-4 \dot{\tilde{\sigma} }^2 g^{\mu \nu }\partial _{\mu }\dot{\tilde{\sigma} }\partial
_{\nu }\dot{\tilde{\sigma} }
\nn\\ &\,&\qquad\qquad\qquad\qquad\quad
+g^{\alpha\beta}\partial_{\alpha }\tilde{\sigma} \partial_{\beta } \tilde{\sigma} g^{\mu \nu }\partial_\mu\partial_\nu \tilde{\sigma} \ddot{\tilde{\sigma}}
\Big)\,,\label{eq:S4-2-220716}
\ea
where we have neglected higher order terms. From Eq. (\ref{eq:S2-2-220716}), we can see that the dispersion relation is simply $F=\omega^2-k^2=0$.
As explained in Sec. \ref{sec:GalileonGenesis-1}, the time scale of the scattering processes we consider will be $\Delta t\ll |t|$
in the regime $|t|\gg 1/M$. Therefore, we can approximately treat the coefficients ${2\tilde{\kappa}_1 (-t) \over \tilde{M}^3}$ and ${\tilde{\kappa}_1 (-t)^2 \over \tilde{M}^4}$ in (\ref{eq:S3-2-220716}) and (\ref{eq:S4-2-220716}) as constants in the calculation of scattering amplitudes. {Again, the energy scale of the scattering processes we consider should satisfy $M> E\sim 1/\Delta t \gg E_{\rm IR} \simeq |t|^{-1}$.}

\subsection{Perturbative unitarity}\label{bH:sec:210806-01}


In this subsection, by using the constraints of perturbative unitarity introduced in Sec. \ref{sec:PU-220717-1}, we revisit the specific ``beyond Horndeski'' action (\ref{eq:210726-1}) in the context of genesis cosmology.
Due to the complexity of the interactions given by Eqs. (\ref{eq:S3-2-220716}) and (\ref{eq:S4-2-220716}), we will calculate only the amplitude for the four-point scattering processes in the following.

\subsubsection{Constraints from four-point scattering}
\begin{figure}[htbp]
	\centering
	\includegraphics[width=0.7\columnwidth]{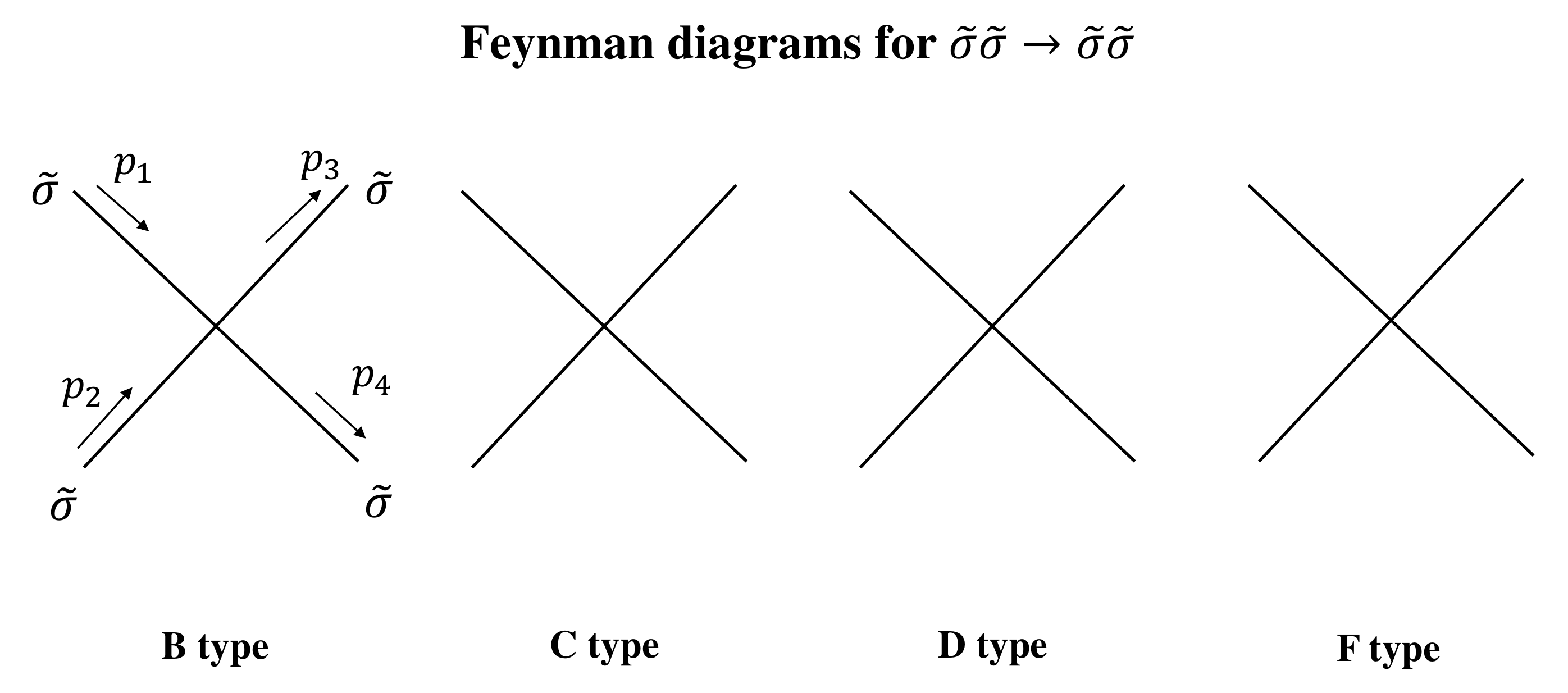}
	\caption{First kind of Feynman diagrams for $\tilde{\sigma}\tilde{\sigma}\to\tilde{\sigma}\tilde{\sigma}$ which only utilize one four-point vertex. The presented four types (B, C, D, F) whose contributions are nonzero correspond to the second, third, fourth and sixth terms in $S^{(4)}_{\tilde{\sigma}}$ in Eq. (\ref{bH:eq:S4}) respectively.}
	\label{bH:FourScatter1}
\end{figure}

\begin{figure}[htbp]
	\centering
	\includegraphics[width=0.65\columnwidth]{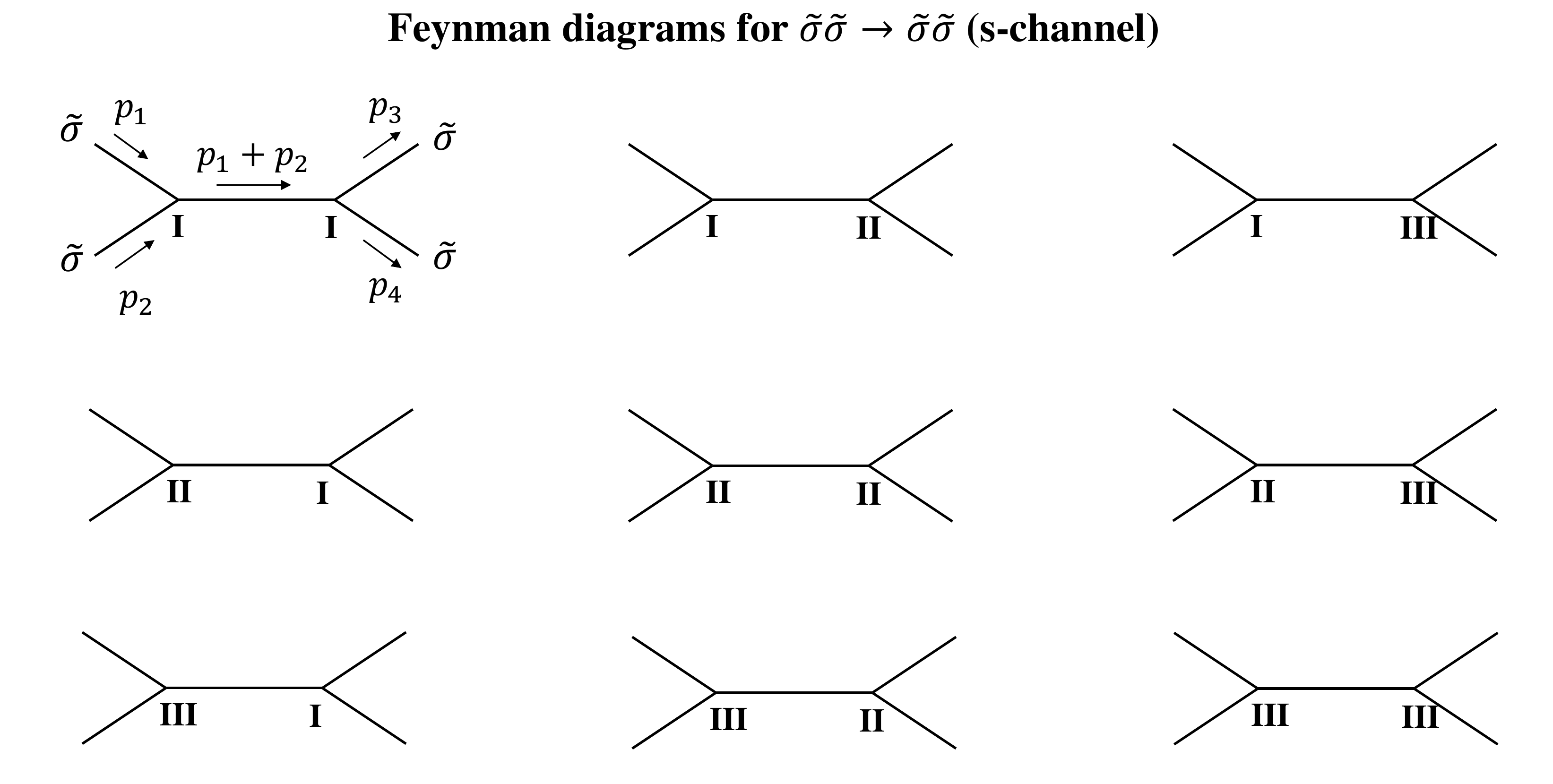}
	\centering
	\includegraphics[width=0.7\columnwidth]{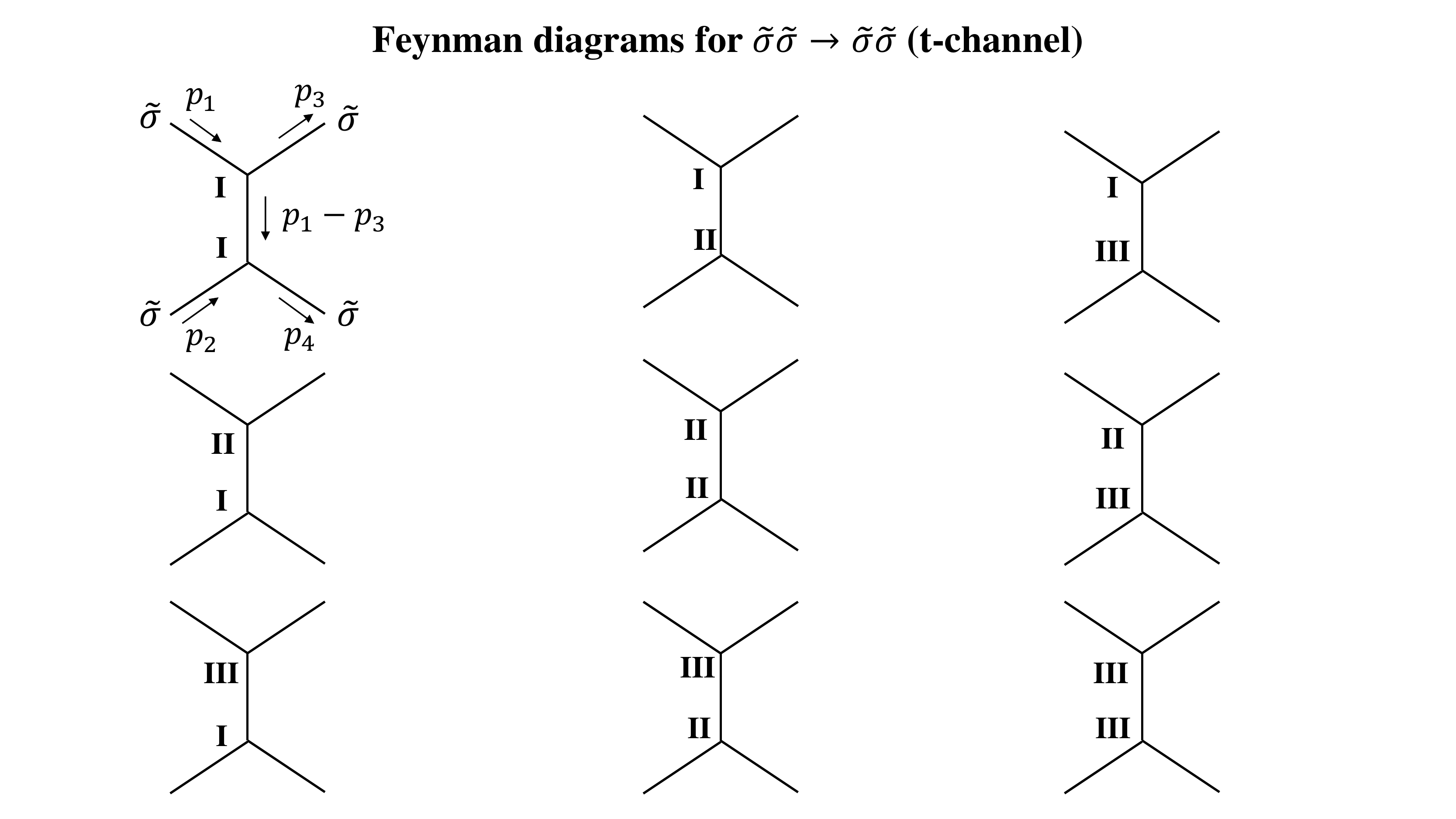}
	\centering
	\includegraphics[width=0.65\columnwidth]{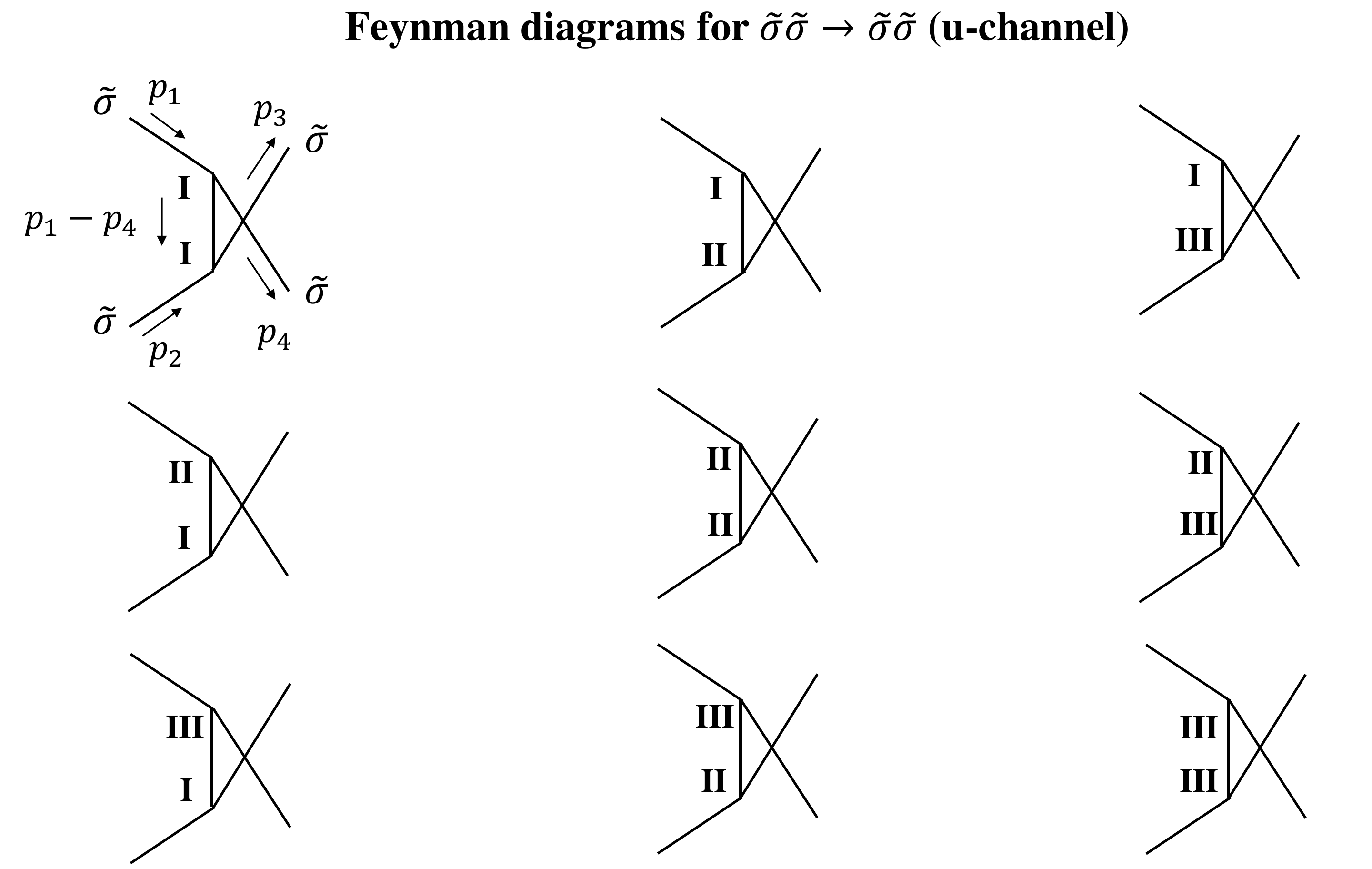}
	\centering
	\caption{Three kinds of Feynman diagrams for $\tilde{\sigma}\tilde{\sigma}\to\tilde{\sigma}\tilde{\sigma}$ which utilize two three-point vertices.}
	\label{bH:FourScatter2}
\end{figure}

The Feynman diagrams for $\tilde{\sigma}\tilde{\sigma}\to\tilde{\sigma}\tilde{\sigma}$ can be categorized into four kinds which are presented in Fig.~\ref{bH:FourScatter1} and Fig.~\ref{bH:FourScatter2} separately.
The amplitudes corresponding to Fig.~\ref{bH:FourScatter1} are
\begin{eqnarray}
i \mathcal{M}_B(\tilde{\sigma}\tilde{\sigma}\to\tilde{\sigma}\tilde{\sigma}) &=& -2i {\tilde{\kappa}_1 (-t)^2 \over \tilde{M}^4} \Bigg\{  -p_{12}\big( p_{14}p_{23} + p_{13}p_{24} \big)  +p_{13}\big( p_{14}(p_{23}+p_{24}) +p_{24}(p_{23}-p_{34}) \big) \nn\\
&&+p_{14}p_{23}\big( p_{24}-p_{34} \big) +p_{12}p_{34}\big( p_{13}+p_{14}+p_{23}+p_{24} \big)  \Bigg\} \,, \nn\\
i \mathcal{M}_C(\tilde{\sigma}\tilde{\sigma}\to\tilde{\sigma}\tilde{\sigma}) &=&  -4i {\tilde{\kappa}_1 (-t)^2 \over \tilde{M}^4} \Bigg\{  p_{14}p_{23}(p_{1t4t} +p_{2t3t})+p_{13}p_{24}(p_{1t3t}+p_{2t4t}) -p_{12}p_{34}(p_{1t2t}+p_{3t4t})  \Bigg\} \,, \nn\\
i \mathcal{M}_D(\tilde{\sigma}\tilde{\sigma}\to\tilde{\sigma}\tilde{\sigma}) &=& -4 i {\tilde{\kappa}_1 (-t)^2 \over \tilde{M}^4}  \Bigg\{  -p_{12} \Big(p_{13} p_{2t4t}+p_{14} p_{2t3t}+p_{24}p_{1t3t}+p_{23}p_{1t4t}\Big) \nn\\
&&+p_{13} \Big( p_{14}(p_{2t3t}+p_{2t4t})+p_{1t2t} p_{34}+p_{23} (p_{1t4t}+p_{2t4t})-p_{2t4t} p_{34} \Big) \nn\\
&&+p_{34} \Big( p_{14}(p_{1t2t}-p_{2t3t})+p_{1t2t} (p_{23}+p_{24})-p_{1t3t} p_{24}-p_{1t4t} p_{23}\Big) \nn\\
&&+p_{24} \Big( p_{14} (p_{1t3t}+p_{2t3t})+p_{23} (p_{1t3t}+p_{1t4t}) \Big) +p_{12} p_{3t4t} \Big( p_{13}+p_{14}+p_{23}+p_{24} \Big)  \Bigg\} \,, \nn\\
i \mathcal{M}_F(\tilde{\sigma}\tilde{\sigma}\to\tilde{\sigma}\tilde{\sigma}) &=& -16i {\tilde{\kappa}_1 (-t)^2 \over \tilde{M}^4} \Big(p_{1t2t}p_{3t4t}\Big) \lf( -p_{12}+p_{13}+p_{14}-p_{34}+p_{23}+p_{24}  \rt) \,,
\end{eqnarray}
where $p_{ij}\equiv p_i \cdot p_j$.
This expression can be written in compact form
\begin{eqnarray}
\mathcal{M}_{4} &=& \mathcal{M}_B +\mathcal{M}_C +\mathcal{M}_D +\mathcal{M}_F \nn\\
&=&  {\tilde{\kappa}_1 (-t)^2 \over \tilde{M}^4}  \frac{1}{2} \lf( \mathsf{s}^3 +\mathsf{t}^3 +\mathsf{u}^3 -2\mathsf{s}^2 \mathsf{s}_\mathsf{t} -2\mathsf{t}^2 \mathsf{t}_\mathsf{t} -2\mathsf{u}^2 \mathsf{u}_\mathsf{t} \rt) \,,
\end{eqnarray}
where the Mandelstam variables
\begin{eqnarray}\label{bH:Mandelstam1}
&\,& \mathsf{s}=-(p_1+p_2)^2=-(p_3+p_4)^2 \,,\nn\\
&\,& \mathsf{t}=-(p_3-p_1)^2=-(p_2-p_4)^2 \,,\nn\\
&\,& \mathsf{u}=-(p_4-p_1)^2=-(p_2-p_3)^2 \,,
\end{eqnarray}
and
\begin{eqnarray}\label{bH:Mandelstam2}
&\,& \mathsf{s}_\mathsf{t}= 2p_{1t2t} = 2p_{3t4t} \,, \quad\quad\,\,\,   \mathsf{s}_\mathsf{s}= -2\vec{p}_1 \cdot\vec{p}_2 = -2\vec{p}_3\cdot \vec{p}_4  \,,\nn\\
&\,& \mathsf{t}_\mathsf{t}= -2p_{1t3t} = -2p_{2t4t} \,, \quad   \mathsf{t}_\mathsf{s}= 2\vec{p}_1\cdot \vec{p}_3 = 2\vec{p}_2 \cdot\vec{p}_4  \,,\nn\\
&\,& \mathsf{u}_\mathsf{t}= -2p_{1t4t} = -2p_{2t3t} \,, \quad   \mathsf{u}_\mathsf{s}= 2\vec{p}_1 \cdot\vec{p}_4 = 2\vec{p}_2 \cdot\vec{p}_3  \,,
\end{eqnarray}
where $p_{itjt}\equiv p_{it} p_{jt}$.

The amplitudes for $\mathsf{s}$-channel in Fig.~\ref{bH:FourScatter2} are
\begin{eqnarray}
\mathcal{M}_{\mathsf{s}\uppercase\expandafter{\romannumeral1}\to\uppercase\expandafter{\romannumeral1}} &=&  {\tilde{\kappa}_1 (-t)^2 \over \tilde{M}^4} p_{12}(p_{1t}+p_{2t}) \bigg( p_{34}(p_{14}+p_{24})(p_{1t}+p_{2t}-p_{3t}) +p_{34}(p_{13}+p_{23})(p_{1t}+p_{2t}-p_{4t}) \nn\\
&&+(p_{13}+p_{23})(p_{14}+p_{24})(p_{3t}+p_{4t}) \bigg) \,,\nn\\
\mathcal{M}_{\mathsf{s}\uppercase\expandafter{\romannumeral1}\to\uppercase\expandafter{\romannumeral2}} &=&  {\tilde{\kappa}_1 (-t)^2 \over \tilde{M}^4}  \bigg(  -2 p_{12}^2 (p_{1t}+p_{2t})(p_{3t}+p_{4t}) \vec{p}_3 \cdot\vec{p}_4  \bigg) \,,\nn\\
\mathcal{M}_{\mathsf{s}\uppercase\expandafter{\romannumeral1}\to\uppercase\expandafter{\romannumeral3}} &=&  {\tilde{\kappa}_1 (-t)^2 \over \tilde{M}^4}  \bigg(  2p_{12}p_{3t4t}(p_{1t}+p_{2t})^2(p_{13}+p_{14}+p_{23}+p_{24}-p_{34})  \bigg) \,,\nn\\
\mathcal{M}_{\mathsf{s}\uppercase\expandafter{\romannumeral2}\to\uppercase\expandafter{\romannumeral1}} &=&  {\tilde{\kappa}_1 (-t)^2 \over \tilde{M}^4} \vec{p}_1\cdot\vec{p}_2(-p_{1t}-p_{2t})\bigg( p_{34}(p_{14}+p_{24})(p_{1t}+p_{2t}-p_{3t}) +p_{34}(p_{13}+p_{23})(p_{1t}+p_{2t}-p_{4t}) \nn\\
&&+(p_{13}+p_{23})(p_{14}+p_{24})(p_{3t}+p_{4t}) \bigg) \,,\nn\\
\mathcal{M}_{\mathsf{s}\uppercase\expandafter{\romannumeral2}\to\uppercase\expandafter{\romannumeral2}} &=&  {\tilde{\kappa}_1 (-t)^2 \over \tilde{M}^4}   \bigg(  2p_{12}(\vec{p}_1\cdot\vec{p}_2)(\vec{p}_3\cdot\vec{p}_4)(p_{1t}+p_{2t})(p_{3t}+p_{4t})  \bigg) \,,\nn\\
\mathcal{M}_{\mathsf{s}\uppercase\expandafter{\romannumeral2}\to\uppercase\expandafter{\romannumeral3}} &=&  {\tilde{\kappa}_1 (-t)^2 \over \tilde{M}^4}  \bigg(  -2(p_{1t}+p_{2t})^2 p_{3t4t} \vec{p}_1\cdot\vec{p}_2 (p_{13}+p_{14}+p_{23}+p_{24}-p_{34})  \bigg) \,,\nn\\
\mathcal{M}_{\mathsf{s}\uppercase\expandafter{\romannumeral3}\to\uppercase\expandafter{\romannumeral1}} &=&  {\tilde{\kappa}_1 (-t)^2 \over \tilde{M}^4}  p_{1t2t}(p_{1t}+p_{2t})\bigg(  p_{34}(p_{14}+p_{24})(p_{1t}+p_{2t}-p_{3t}) +p_{34}(p_{13}+p_{23})(p_{1t}+p_{2t}-p_{4t}) \nn\\ &&+(p_{13}+p_{23})(p_{14}+p_{24})(p_{3t}+p_{4t})  \bigg) \,,\nn\\
\mathcal{M}_{\mathsf{s}\uppercase\expandafter{\romannumeral3}\to\uppercase\expandafter{\romannumeral2}} &=&  {\tilde{\kappa}_1 (-t)^2 \over \tilde{M}^4}  \bigg(  -2p_{12}p_{1t2t} \vec{p}_3\cdot\vec{p}_4(p_{1t}+p_{2t})(p_{3t}+p_{4t})  \bigg) \,,\nn\\
\mathcal{M}_{\mathsf{s}\uppercase\expandafter{\romannumeral3}\to\uppercase\expandafter{\romannumeral3}} &=&  {\tilde{\kappa}_1 (-t)^2 \over \tilde{M}^4}   \bigg(  2p_{1t2t}p_{3t4t}(p_{1t}+p_{2t})^2(p_{13}+p_{14}+p_{23}+p_{24}-p_{34})  \bigg) \,.
\end{eqnarray}
This expression can be written in compact form in terms of Eq.~(\ref{bH:Mandelstam1}) and Eq.~(\ref{bH:Mandelstam2}),
\begin{eqnarray}
\mathcal{M}_{\mathsf{s}} &=& \mathcal{M}_{\mathsf{s}\uppercase\expandafter{\romannumeral1}\to\uppercase\expandafter{\romannumeral1}} +\mathcal{M}_{\mathsf{s}\uppercase\expandafter{\romannumeral1}\to\uppercase\expandafter{\romannumeral2}} +\mathcal{M}_{\mathsf{s}\uppercase\expandafter{\romannumeral1}\to\uppercase\expandafter{\romannumeral3}} +\mathcal{M}_{\mathsf{s}\uppercase\expandafter{\romannumeral2}\to\uppercase\expandafter{\romannumeral1}} +\mathcal{M}_{\mathsf{s}\uppercase\expandafter{\romannumeral2}\to\uppercase\expandafter{\romannumeral2}} +\mathcal{M}_{\mathsf{s}\uppercase\expandafter{\romannumeral2}\to\uppercase\expandafter{\romannumeral3}} \nn\\ &&+\mathcal{M}_{\mathsf{s}\uppercase\expandafter{\romannumeral3}\to\uppercase\expandafter{\romannumeral1}}  +\mathcal{M}_{\mathsf{s}\uppercase\expandafter{\romannumeral3}\to\uppercase\expandafter{\romannumeral2}}  +\mathcal{M}_{\mathsf{s}\uppercase\expandafter{\romannumeral3}\to\uppercase\expandafter{\romannumeral3}} \nn\\
&=& {\tilde{\kappa}_1 (-t)^2 \over \tilde{M}^4} \frac{1}{4}{\mathsf{s}(\mathsf{s}-\mathsf{s})}\Bigg\{ \mathsf{s}(p_{1t}^2+p_{2t}^2) -\mathsf{s}_\mathsf{t}(p_{1t}^2+p_{2t}^2-\mathsf{s}+\mathsf{s}_\mathsf{t}) +\mathsf{s}_\mathsf{s}(\mathsf{t}_\mathsf{t}+\mathsf{u}_\mathsf{t})  \Bigg\} \,.
\end{eqnarray}
Similarly, the amplitudes for $\mathsf{u}$-channel and $\mathsf{t}$-channel in Fig.~\ref{bH:FourScatter2} are
\begin{eqnarray}
\mathcal{M}_{\mathsf{t}} &=& \mathcal{M}_{\mathsf{t}\uppercase\expandafter{\romannumeral1}\to\uppercase\expandafter{\romannumeral1}} +\mathcal{M}_{\mathsf{t}\uppercase\expandafter{\romannumeral1}\to\uppercase\expandafter{\romannumeral2}} +\mathcal{M}_{\mathsf{t}\uppercase\expandafter{\romannumeral1}\to\uppercase\expandafter{\romannumeral3}} +\mathcal{M}_{\mathsf{t}\uppercase\expandafter{\romannumeral2}\to\uppercase\expandafter{\romannumeral1}} +\mathcal{M}_{\mathsf{t}\uppercase\expandafter{\romannumeral2}\to\uppercase\expandafter{\romannumeral2}} +\mathcal{M}_{\mathsf{t}\uppercase\expandafter{\romannumeral2}\to\uppercase\expandafter{\romannumeral3}} \nn\\ &&+\mathcal{M}_{\mathsf{t}\uppercase\expandafter{\romannumeral3}\to\uppercase\expandafter{\romannumeral1}}  +\mathcal{M}_{\mathsf{t}\uppercase\expandafter{\romannumeral3}\to\uppercase\expandafter{\romannumeral2}}  +\mathcal{M}_{\mathsf{t}\uppercase\expandafter{\romannumeral3}\to\uppercase\expandafter{\romannumeral3}} \nn\\
&=&  {\tilde{\kappa}_1 (-t)^2 \over \tilde{M}^4} \frac{1}{4}{\mathsf{t}(\mathsf{t}-\mathsf{t})}\Bigg\{ \mathsf{t}(p_{1t}^2+p_{3t}^2) -\mathsf{t}_{\mathsf{t}}(p_{1t}^2+p_{3t}^2-\mathsf{t}) -\mathsf{s}_{\mathsf{t}}^2 +\mathsf{t}_\mathsf{s}(\mathsf{s}_\mathsf{t}+\mathsf{u}_\mathsf{t})  \Bigg\} \,,
\end{eqnarray}
and
\begin{eqnarray}
\mathcal{M}_{\mathsf{u}} &=& \mathcal{M}_{\mathsf{u}\uppercase\expandafter{\romannumeral1}\to\uppercase\expandafter{\romannumeral1}} +\mathcal{M}_{\mathsf{u}\uppercase\expandafter{\romannumeral1}\to\uppercase\expandafter{\romannumeral2}} +\mathcal{M}_{\mathsf{u}\uppercase\expandafter{\romannumeral1}\to\uppercase\expandafter{\romannumeral3}} +\mathcal{M}_{\mathsf{u}\uppercase\expandafter{\romannumeral2}\to\uppercase\expandafter{\romannumeral1}} +\mathcal{M}_{\mathsf{u}\uppercase\expandafter{\romannumeral2}\to\uppercase\expandafter{\romannumeral2}} +\mathcal{M}_{\mathsf{u}\uppercase\expandafter{\romannumeral2}\to\uppercase\expandafter{\romannumeral3}} \nn\\ &&+\mathcal{M}_{\mathsf{u}\uppercase\expandafter{\romannumeral3}\to\uppercase\expandafter{\romannumeral1}}  +\mathcal{M}_{\mathsf{u}\uppercase\expandafter{\romannumeral3}\to\uppercase\expandafter{\romannumeral2}}  +\mathcal{M}_{\mathsf{u}\uppercase\expandafter{\romannumeral3}\to\uppercase\expandafter{\romannumeral3}} \nn\\
&=&  {\tilde{\kappa}_1 (-t)^2 \over \tilde{M}^4} \frac{1}{4}{\mathsf{u}(\mathsf{u}-\mathsf{u})}\Bigg\{ \mathsf{u}(p_{1t}^2+p_{4t}^2) -\mathsf{u}_{\mathsf{t}}(p_{1t}^2+p_{4t}^2-\mathsf{u}) -\mathsf{s}_{\mathsf{t}}^2 +\mathsf{u}_{\mathsf{s}}(\mathsf{s}_{\mathsf{t}}+\mathsf{t}_{\mathsf{t}})  \Bigg\} \,.
\end{eqnarray}
The total amplitude is
\begin{eqnarray}
\mathcal{M}_{\rm total} = \mathcal{M}_{4}+\mathcal{M}_{\mathsf{s}}+\mathcal{M}_{\mathsf{t}}+\mathcal{M}_{\mathsf{u}} \,.
\end{eqnarray}

Therefore, the stable NEC violation in the context of ``beyond Horndeski'' genesis, which is constructed by the specific action (\ref{eq:210726-1}) with $F$ given by Eq. (\ref{eq:FX0715}), cannot satisfy the requirement (\ref{constrain}) of UV completion.
In the EFT regime, by using the constraint (\ref{constrain-EFT-1}) of perturbative unitarity, we obtain approximately
\be \sqrt{\mathsf{s}}\lesssim (M/M_{\rm P})^{1/3} |t|^{-1}\sim (M/M_{\rm P})^{1/3} \tilde{M}\ll M \,,
\ee
where we have disregarded the coefficient.
Namely, the perturbative unitarity is violated at a scale $\sqrt{s} \sim (M/M_{\rm P})^{1/3} |t|^{-1}\sim (M/M_{\rm P})^{1/3} \tilde{M}$ $\ll M$.
Given the IR cut-off scale $E_{\rm IR}$ of the scattering processes we considered, the tree-level perturbative unitarity is already violated.
This result is actually consistent with the results of Secs. \ref{sec:4GPUB-1} and \ref{sec:5GPUB-1}, despite the differences in coefficients.

Consequently, although we are able to realize fully stable NEC violation with the ``beyond Horndeski'' theory, the constraints of tree-level perturbative unitarity imply that we may need some unknown new physics below the cut-off scale $M$ in the EFT other than that represented by the ``beyond Horndeski'' operators in Eq. (\ref{eq:220122L-1}) or (\ref{eq:210726-1}).



\section{Summary and outlook}\label{Summary}

The NEC is a crucial and quite robust condition in gravity and cosmology. The explorations of a fully stable NEC violation have made some significant progress. However, it is still an open question whether some fundamental properties of UV-complete theories or the consistency requirements of EFT forbid a violation of the NEC.
The constraints of perturbative unitarity could provide us with some novel insights into the EFT of a fully stable NEC violation.

In this paper, we investigated the tree-level perturbative unitarity for stable NEC violations in the contexts of both Galileon genesis and ``beyond Horndeski'' genesis cosmology, in which the universe is asymptotically Minkowskian in the infinite past. The calculations of scattering amplitudes could be simplified by the genesis solution provided we consider only the leading order interacting terms.
It is found that the tree-level perturbative unitarity gets broken at an energy scale $\sqrt{\mathsf{s}}\sim |t|^{-1}\ll M$ in both Galileon genesis and ``beyond Horndeski'' genesis, where $M$ is the cut-off scale of the EFT action, $t$ is the cosmological time. Therefore, the constraints of perturbative unitarity imply that the earlier era (i.e., the larger of $|t|$) we go into the genesis phase, the more urgent we may need the unknown new physics at a lower scale other than that represented by the ``beyond Horndeski'' operators.

In the calculations of the scattering amplitudes, we have assumed that the sound speed squared $c_s^2=1$ during the NEC-violating phase for simplicity. Additionally, the models of genesis cosmology we considered are constructed by some specific Galileon and ``beyond Horndeski'' theories, which are representative to some extent. Whether our conclusion remains unchanged for a general $c_s^2$ and other more complicated ``beyond Horndeski'' theories (or a different construction of the stable genesis model as discussed in \cite{Cai:2017tku}) requires further investigations. It would also be interesting to see whether the required unknown new physics indicated by perturbative unitarity can be represented by those higher order DHOST operators or some modified dispersion relations. Our study might be extended to the context of bouncing cosmology as well. Furthermore, taking into account the constraints from cosmological observations will place tighter constraints on the EFT of a fully stable NEC violation.

\acknowledgments

We would like to thank Yun-Song Piao, Shuang-Yong Zhou, Toshifumi Noumi, Gen Ye, Yunlong Zheng, Mian Zhu and Chao Chen for helpful discussions.
The work of Cai is supported in part by the National Natural Science Foundation of China (Grant No. 11905224), the China Postdoctoral Science Foundation (Grant No. 2021M692942) and Zhengzhou University (Grant No. 32340282).
The work of Xu is supported in part by National Natural Science Foundation of China under Grant Nos. 12105247 and 12047545, the China Postdoctoral Science Foundation under Grant No. 2021M702957.
The work of Zhao is supported by Jefferson Science Associates, LLC under  U.S. DOE Contract \# DE-AC05-06OR23177 and by U.S. DOE Grant \# DE-FG02-97ER41028.
The work of Zhou  is supported in part by the Swedish Research Council under grants number 2015-05333 and 2018-03803.
We acknowledge the use of the computing server {\it Arena317}@ZZU.

\appendix


\section{The EFT of nonsingular cosmology} \label{sec:app-EFT-1}

The approach of the effective field theory (EFT) is powerful in investigating inflation \cite{Cheung:2007st}, dark energy \cite{Gubitosi:2012hu,Gleyzes:2013ooa,Piazza:2013coa} as well as nonsingular cosmology \cite{Cai:2016thi,Creminelli:2016zwa,Cai:2017tku,Cai:2017dyi}.
We work with the $3 + 1$ decomposed metric $d s^{2}=-N^{2} d t^{2}+h_{i j}\left(d x^{i}+N^{i} d t\right)\left(d x^{j}+N^{j} d t\right)$, where $h_{\mu \nu} = g_{\mu \nu}+n_{\mu} n_{\nu}$ is the induced metric, $n^\mu$ is the unit normal vector of the constant time hypersurfaces, $N$ and $N^i$ are the lapse function and the shift vector, respectively.
In the unitary gauge, the EFT action that is able to realize a fully stable NEC violation can be written as
\ba
S&=&\int
d^4x\sqrt{-g}\Big[ {M_{\rm P}^2\over2} f(t)R-\Lambda(t)-c(t)g^{00}
\nn\\
&\,&\qquad\qquad\quad +{M_2^4(t)\over2}(\delta g^{00})^2-{m_3^3(t)\over2}\delta
K\delta g^{00} -m_4^2(t)\lf( \delta K^2-\delta K_{\mu\nu}\delta
K^{\mu\nu} \rt)
\nn\\
&\,&\qquad\qquad\quad  + {\tilde{m}_4^2(t)\over
	2}R^{(3)}\delta g^{00}\Big]
\label{action01}
\ea
up to quadratic order \cite{Cai:2016thi,Creminelli:2016zwa,Cai:2017tku,Cai:2017dyi} (see also \cite{Cheung:2007st,Gubitosi:2012hu,Gleyzes:2013ooa,Piazza:2013coa}),
where $R^{(3)}$ is the induced 3-dimensional Ricci scalar, $K_{\mu\nu}$ is the extrinsic curvature, $\delta g^{00}=g^{00}+1$, $\delta K_{\mu\nu}=K_{\mu\nu}-h_{\mu\nu}H$, $H$ is the Hubble parameter.
We have disregarded those higher-order spatial derivatives and the action of matter sector in (\ref{action01}).

In the cosmological context where there is an evolving scalar field $\phi$, the constant time hypersurfaces can be set as the uniform scalar field hypersurfaces. As a result, we have $n_\mu=-\phi_\mu/\sqrt{-X}$, where we have defined $X=\phi_{\mu} \phi^{\mu}$, $\phi_{\mu}=\nabla_{\mu} \phi$ and $\phi^{\mu}=\nabla^{\mu} \phi$ for convenience. Using $K_{\mu\nu}=h_\mu^\sigma\nabla_\sigma n_\nu$ and the  Gauss-Codazzi relation, we can obtain the corresponding covariant action of (\ref{action01}) in principle, see e.g. \cite{Gleyzes:2013ooa,Gleyzes:2014rba}.
In fact, the EFT action (\ref{action01}) is able to specify a variety of theories of gravity by different choices of the time-dependent coefficiens $f$, $\Lambda$, $c$, $M_2^4$, $m_3^3$, $m_4^2$ and $\tilde{m}_4^2$. For example, the Horndeski theory \cite{Horndeski:1974wa,Deffayet:2011gz,Kobayashi:2011nu} and the ``beyond Horndeski'' (GLPV) theory \cite{Gleyzes:2014dya,Gleyzes:2014qga} correspond to $m_4^2= {\tilde m}_4^2$ and $m_4^2\neq {\tilde m}_4^2$, respectively.

However, in order to cover more general degenerate higher-order scalar-tensor (DHOST) theories \cite{Langlois:2015cwa} (see \cite{Langlois:2018dxi} for a review), the EFT action (\ref{action01}) has to incorporate additional operators $\delta K \dot{\delta g^{00}}$, $(\dot{\delta g^{00}})^2$ and $(\partial_i{\delta g^{00}})^2$ \cite{Gleyzes:2014rba,Langlois:2017mxy,Langlois:2017mdk}, where there is still no Ostrogradski instability.
Following the convention of \cite{Langlois:2017mxy,Langlois:2017mdk,Langlois:2018dxi}, the action of all quadratic and cubic DHOST theories can be written as
\ba \label{eq:21080102}
&\,&S^{\text {quad}}=\int d^{3} x d t a^{3} \frac{{\cal M}^{2}}{2}\left\{\delta K_{i j} \delta K^{i j}-\left(1+\frac{2}{3} \alpha_{\mathrm{L}}\right) \delta K^{2}+\left(1+\alpha_{\mathrm{T}}\right)\left(\delta_1 R^{(3)} \frac{\delta \sqrt{h}}{a^{3}}+\delta_{2} R^{(3)} \right)\right. \nn
\\
&\,&\qquad\qquad\qquad\qquad\qquad\quad
+H^{2} \alpha_{\mathrm{K}} \delta N^{2}+4 H \alpha_{\mathrm{B}} \delta K \delta N+\left(1+\alpha_{\mathrm{H}}\right) \delta_1 R^{(3)} \delta N
\nn\\&\,&\qquad\qquad\qquad\qquad\qquad\quad
\lf.
+4 \beta_{1} \delta K \delta \dot{N}+\beta_{2} \delta \dot{N}^{2}+\frac{\beta_{3}}{a^{2}}\left(\partial_{i} \delta N\right)^{2}\right\}
\ea
up to quadratic order in the unitary gauge,
where $\delta N \equiv 1-N=\delta g^{00} / 2$. Note that the contribution from the first line of (\ref{action01}) at quadratic order is also included in (\ref{eq:21080102}). The relations between ${\cal M}^{2}$, $\alpha_i$ and the coefficients in (\ref{eq:21080102}) can be find in \cite{Gleyzes:2014rba}. The covariant scalar-tensor theories can be mapped to (\ref{eq:21080102}) by using the relations provided in \cite{Langlois:2017mxy}.
Particularly, for the Horndeski theory and the ``beyond Horndeski'' (GLPV) theory, which belong to the subclass Ia of the DHOST theory, we will find $\beta_i=0$ in (\ref{eq:21080102}).


\bibliography{Setting/ref}
\bibliographystyle{Setting/jhep}

\end{document}